\documentclass[12pt,a4paper,english,superscriptaddress,aps,nofootinbib]{revtex4}
\usepackage[utf8]{inputenc}
\usepackage[T1]{fontenc}
\usepackage{amsmath}
\usepackage{amssymb}
\usepackage{graphicx}
\makeatletter
\usepackage{babel}
\usepackage[active]{srcltx}
\usepackage{graphicx,color}
\usepackage{subfigure}
\usepackage{changebar}
\usepackage{upgreek}
\usepackage{hyperref}
\hypersetup{colorlinks=true,urlcolor= blue,citecolor=blue,linkcolor= blue}
\usepackage{braket}
\usepackage{dsfont}
\usepackage{mathtools}
\usepackage{slashed}
\usepackage{empheq}
\usepackage{tikz}
\usepackage{tikz-feynman}
\usepackage{multirow}
\usetikzlibrary{decorations.pathmorphing}

\usepackage{tikz}
\usepackage{tikz-feynman}
\tikzfeynmanset{compat=1.1.0}

\usepackage{xcolor}

\usepackage{graphicx}
\usepackage{amsfonts}
\usepackage{amsmath}
\newcommand{\CP}{\mathbb{C}\mathrm{P}}

\newcommand{\be}{\begin{equation}}
	\newcommand{\ee}{\end{equation}}
\newcommand{\bea}{\begin{eqnarray}}
	\newcommand{\eea}{\end{eqnarray}}

\begin{document}

\title{BPS vortex from nonpolynomial scalar QED in a $\CP^1$--Maxwell theory}

\author{F. C. E. Lima}
\email[]{E-mail: cleiton.estevao@fisica.ufc.br; cleiton.estevao@ufabc.edu.br}
\affiliation{Centro de Matématica, Computação e Cognição (CMCC), Universidade Federal do ABC (UFABC), Av. dos Estados 5001, CEP 09210-580, Santo Andr\'{e}, S\~{a}o Paulo, Brazil.}

\begin{abstract}
\noindent \textbf{Abstract:} We investigate a generalized gauged $\mathbb{CP}^{1}$-Maxwell theory in which the electromagnetic sector acquires a field-dependent magnetic permeability generated dynamically through fermionic vacuum polarization. We couple a Dirac fermion with an effective mass to a gauged $\mathbb{CP}^{1}$ sigma model whose target space carries the Fubini–Study metric. Integrating out the Dirac fermion at one loop induces a logarithmic correction to the Maxwell kinetic term. This dependence is retained in the planar effective theory after dimensional reduction. Within this framework, one builds a generalized $\mathbb{CP}^{1}$-Maxwell model by admitting Bogomol'nyi-Prasad-Sommerfield (BPS) configurations. Taking this into account, we solved the self-dual equations that describe vortex-like solutions with quantized magnetic flux.
\end{abstract}

\maketitle
\newpage

\section{Introduction}

The first discussions of vortex solutions emerged in condensed matter physics in 1957 \cite{Abrikosov1,Abrikosov2}. In this work,  Abrikosov demonstrated that type-II superconductors admit configurations in which the magnetic field penetrates the material in the form of quantized flux lines, now known as Abrikosov vortices. These results appear within the framework of the Ginzburg-Landau theory of superconductivity \cite{Ginzburg} and established the physical basis for understanding topological defects in two-dimensional systems with spontaneous symmetry breaking \cite{Kibble1,Kibble2,Vilenkin}. Subsequently, Nielsen and Olesen found analogous solutions in the relativistic context by adopting an Abelian-Higgs model \cite{Nielsen}. In their classic work ``\textit{Vortex-Line Models for Dual Strings}'', the authors showed that a gauge theory with spontaneous symmetry breaking admits static solutions with finite energy per unit length, corresponding to tubes of quantized magnetic flux \cite{Nielsen}. These configurations naturally possess a topological character associated with the fundamental homotopy group of the vacuum manifold. Consequently, vortex solutions have come to play a central role in several areas of theoretical physics, including the study of topological defects \cite{Bazeia2,Bazeia3,Lima2,Lima3,Petrov}, cosmic strings \cite{Linet,Kim,Fedderke}, and confinement mechanisms in gauge theories \cite{Hayata}.

An important advance in the understanding of these solutions occurred a few years later with the introduction of the Bogomol'nyi-Prasad-Sommerfield (BPS) framework \cite{Bogomolnyi,Prasad}. In 1976, Bogomol'nyi demonstrated that, for a specific value of the coupling ratio in the Abelian-Higgs model, the static energy configurations enable the emergence of a topological lower bound. In this special case, the equations of motion reduce to first-order equations, known as the BPS equations (or self-dual equations), whose solutions saturate the minimum-energy bound \cite{Bogomolnyi}. This result shows that topological defects may, a priori, be interpreted as stable topological configurations whose energy is proportional to the topological charge of the field \cite{Bogomolnyi}. Concomitantly, the work of Prasad and Sommerfield on monopoles in non-Abelian gauge theories reveals an analogous mathematical structure \cite{Prasad}, consolidating what is now referred to as the BPS bound \cite{Bogomolnyi,Prasad}. Since these pioneering developments, the BPS formulation has become a central tool in the study of topological defects in field theory  \cite{Martinec,Alonso2,Lu2025,Kim2025}, providing a simpler and more elegant description of vortex-like solutions \cite{Andrade22} and establishing deep connections with geometric and topological structures in gauge theories \cite{Cunha}.

Another direction explored in the literature concerns the generalization of scalar electrodynamics (QED) in the study of vortex solutions \cite{Petrov,IAndrade11}. Such extensions arise naturally in the context of effective field theories and in analyses of nonlinear realizations of spontaneously broken symmetries \cite{Weinberg1968,Coleman1969,Callan1969}. In these works, one modifies the scalar-field potential or the kinetic sector, incorporating non-polynomial dependencies, allowing for a more general description of the dynamics of the effective Higgs field \cite{Weinberg1968,Coleman1969,Callan1969}. Within scalar QED, we employed these generalizations to investigate the modifications of the topological solutions in a medium characterized by a nonlinear magnetic permeability that emerges from the $\CP^1$-Maxwell theory. This hypothesis opens the possibility of a broader class of models and vortex configurations in which the presence of non-polynomial dielectric functions may significantly affect relevant physical properties of the solutions, such as their energy, spatial profiles, and the conditions required to saturate the BPS bound. Considering these aspects, we will implement this approach for the first time within the framework of the $\CP^1$-Maxwell theory.

Naturally, several models have been gauged to investigate the emergence of topological structures. Among these, nonlinear sigma models (NLSMs) occupy a prominent position \cite{Rajaraman}. Within the wide class of NLSMs discussed in the literature, particular attention has been devoted to the O(3)-sigma model and its equivalent formulation, i.e., the $\CP^{1}$ model \cite{Belavin}, mainly due to their connection with ferromagnetic systems and their applications in condensed matter physics \cite{Polyakov2}. Generally speaking, the $\CP^1$ model emerged in the works of Belavin and Polyakov \cite{Belavin,Polyakov,Polyakov2}, who demonstrated that the two-dimensional O(3)-sigma model admits nontrivial classical solutions with finite energy, known as instantons or topological solitons. These configurations are characterized by a topological charge associated with the homotopy group $\pi_{2}(\mathbb{S}^{2})$, reflecting the geometric structure of the target manifold. Consequently, one notes that the $\CP^{1}$ model provides a mathematically equivalent formulation, allowing for a more convenient description of these solutions in terms of complex fields subject to a local gauge symmetry \cite{Polyakov2}. These results had a profound impact, establishing deep connections between NLSMs, geometric structures, and analytical techniques resembling the BPS approach. As a result, they greatly expanded the understanding of the topological and dynamical properties of solitons across various physical contexts. Despite existing research, there is still a gap in the literature regarding the BPS structure of the $\CP^1$ model when combined with contributions from nonpolynomial scalar QED. Driven by this observation, we will investigate the BPS vortex in nonpolynomial scalar QED within the context of $\CP^1$-Maxwell theory.

Our primary purpose is to investigate the topological solutions of the generalized $\CP^1$–Maxwell theory, where the electromagnetic sector generates a field-dependent magnetic permeability due to the fermionic vacuum polarization. Starting from a microscopic framework in which a Dirac fermion possesses an effective mass depending on the $\CP^1$ field, we show that one-loop quantum corrections induce a non-polynomial contribution to the Maxwell sector. After dimensional reduction to $(2+1)$-dimensions, this mechanism gives rise to a logarithmic dielectric function that modifies the gauge dynamics at the effective level. Within this effective description, we will build a generalized $\CP^1$-Maxwell theory. The relevance of this study lies in establishing a direct connection between quantum vacuum polarization effects, the geometry of the $\CP^1$ target space (concerning the Fubini-Study metric, see Ref. \cite{Manton2}), and physical properties of the topological defects.

We organized this article into seven sections. In Section II, one presents the formulation of the gauged $\CP^1$-Maxwell model, introducing the parametrization of the complex scalar field, the effective action on the target space, and the equations of motion describing the nonlinear coupling between the scalar sector and the gauge field. In Section III, we investigate the mechanism responsible for the dynamical generation of a field-dependent magnetic permeability through fermionic vacuum polarization, explicitly showing how one-loop corrections induce an effective term of the form $G(|u|)F_{\mu\nu}F^{\mu\nu}$. Subsequently, one performs the dimensional reduction to $(2+1)$ dimensions, obtaining the effective form for the electromagnetic sector. In Section IV, one studies the BPS property of the generalized model. In Section V, we examine the physical properties of the magnetized topological defects, discussing the asymptotic behavior of the solutions and their linear stability. Subsequently, in Section VI, we solve the BPS equations numerically in several physical regimes. Lastly, we announced our findings and their physical implications in Section VII.

\section{Formulation of the gauged $\CP^1$ model}

The $\mathbb{CP}^1$ model occupies a central role in theoretical physics, as it provides an exceptionally rich conceptual laboratory for understanding non-perturbative phenomena in quantum field theory \cite{Ward,Leese,Nahum,Buccio}. In two dimensions, the $\mathbb{CP}^1$ model is equivalent to the nonlinear $\mathrm{O}(3)$ sigma model \cite{Polyakov}, which admits topological solutions whose existence is associated with the geometric structure of the target space and to the homotopy classification $\pi_2(\CP^1)=\mathbb{Z}$. Following the seminal work of Polyakov, and in particular, Belavin and Polyakov \cite{Belavin}, the $\mathbb{CP}^1$ model became a paradigm for the study of solitons and instantons \cite{Lian,Ward,Leese,Nahum,Buccio}. Moreover, it emerges as an effective low-energy limit of gauge theories, describes collective excitations in condensed matter systems, e.g., ferromagnets and quantum Hall systems, and serves as a testing ground for supersymmetric extensions, whose non-perturbative features can be systematically examined \cite{Polyakov,Belavin,Rajaraman}. Consequently, the $\mathbb{CP}^1$ model constitutes a fundamental bridge between topology, geometry, and non-perturbative dynamics across different areas of theoretical physics \cite{Polyakov,Belavin,Rajaraman}.

To define the generalized $\mathbb{CP}^1$–Maxwell model, let us start consider the action\footnote{For the development of this work, we adopt the metric signature $g_{\mu\nu}$=diag$(+,-,-)$.}
\begin{align}
    \label{Eq1}
    \mathcal{S}=\int\,d^3x\,\left[\frac{1}{2}(D_\mu \mathbf{n})^2-\frac{G}{4}F_{\mu\nu}F^{\mu\nu}+\tilde{\mathcal{V}}(\vert\mathbf{n}\vert)\right]
\end{align}
where $\mathbf{n}$ is the sigma field, i.e.,
\begin{align}\label{Eq2}
    \mathbf{n}=
    (n^1,n^2,n^3) \qquad \mathrm{with} \qquad \mathbf{n}\cdot\mathbf{n}=1.
\end{align}
Here, it is essential to highlight that $\tilde{V}(\mathbf{n})$ represents the interaction term, while
$F_{\mu\nu}=\partial_\mu A_\nu-\partial_\nu A_\mu$, where $A_\mu$ is  the gauge field. Furthermore, $G$ is the magnetic permeability function.
Indeed, several works have investigated BPS vortex solutions with magnetic permeability; see, for instance, Refs. \cite{Andrade,Bazeia2,Bazeia3,Lima2}. Generalized models that incorporate a magnetic permeability function are particularly relevant because the permeability can effectively extend the symmetry, allowing for the emergence of new topological properties, for instance, the formation of ring-like vortices \cite{Lima3}.

To advance with our analysis, let us adopt a parametrization in terms of the generators of the internal space, i.e., $\sigma^{i}$, such that $n^{a} = z^{\dagger} \sigma^{a} z$ where $z$ is the $\mathbb{CP}^{1}$ field, viz.,
\begin{align}\label{Eq3}
    z=\begin{pmatrix} z_1 \\ z_2\end{pmatrix} \qquad \mathrm{with} \qquad z^\dagger z = 1.
\end{align}
Within this framework, the generators $\sigma^i$ are the Pauli matrices, i.e.,
\begin{align}\label{Eq4}
    \sigma^1=\begin{pmatrix} 0 & 1 \\ 1 & 0 \end{pmatrix}, \qquad \sigma^2=\begin{pmatrix} 0 & -i \\ i & 0 \end{pmatrix}, \qquad \sigma^3=\begin{pmatrix} 1 & 0 \\ 0 & -1\end{pmatrix},
\end{align}
which form a basis of the internal $\mathrm{SU}(2)$ algebra. 

By employing the parametrization $n^{j}=z^{\dagger}\sigma^{j}z$ terms of the internal-space generators \eqref{Eq4}, one obtains the effective action
\begin{align}\label{Eq5}
    \mathcal{S}^{\CP^1}=\int\, d^3x\,\left[\frac{1}{2}(D_\mu z)^{\dagger}(1-z\,z^\dagger)(D^\mu z)-\frac{G}{4}F_{\mu\nu}F^{\mu\nu}-\tilde{\mathcal{V}}\right].
\end{align}

Let us now define the $\mathbb{CP}^1$ field \cite{Polyakov} as
\begin{align}\label{Eq6}
    z=\frac{1}{\sqrt{1+\vert u\vert^2}}\begin{pmatrix} 1 \\ u \end{pmatrix},
\end{align}
which allows us to write the effective $\mathbb{CP}^1$ action in the target space as
\begin{align}
    \label{Eq7} \mathcal{S}^{\mathrm{CP}^1}_{\mathrm{eff}}=\int\, d^3x\,\left[\frac{2}{(1+\vert u\vert^2)^2}(D_\mu u)^{*}(D^\mu u)-\frac{G(\vert u\vert)}{4}F_{\mu\nu}F^{\mu\nu}-\tilde{\mathcal{V}}(\vert u\vert)\right].
\end{align}
The covariant derivative is $D_\mu u=\partial_\mu u-i A_\mu u$, where $\partial_\mu u$ represents the partial derivative of $u$ concerning the coordinate $x^\mu$, and $A_\mu$ is a gauge field\footnote{In this work, one adopts natural units, i.e., $e=\hbar=c=1$.}.

Within this framework, the equations of motion arising from the variation of the action concerning the $\CP^1$ and the gauge fields are given, respectively, by
\begin{align}\label{Eq8}
    D_\mu\left(\frac{2}{(1+| u|^2)^2}D^\mu u\right)+\frac{4u}{(1+|u|^2)^3}(D_\mu u)^*\,(D^\mu u)+\frac{G'(|u|)}{4}\frac{\partial |u|}{\partial\overline{u}}F_{\mu\nu}F^{\mu\nu}+\frac{\partial\tilde{\mathcal{V}}}{\partial\overline{u}}=0
\end{align}
and
\begin{align}
        \label{Eq9} J^\mu=\partial_\nu(GF^{\nu\mu})=i\Omega[u^*D^\mu u-u(D^\mu u)^*],
\end{align}
where $\Omega=2/(1+|u|^2)^2$. Note that the equations of motion derived from the effective action of the generalized $\CP^1$-Maxwell model encode, in a highly nonlinear manner, the coupling between the geometry of the target space and the gauge sector. The scalar field equation involves the geometric factor $\Omega$, which corresponds to the Fubini-Study metric on the sphere $S^2$, indicating that the dynamics of the complex field $u$ takes place on a curved target space. In turn, the modified Maxwell equation incorporates the dielectric function $G(\lvert u\rvert)$, causing the propagation of the electromagnetic field to depend explicitly on the local configuration of the scalar field. As a consequence, the system exhibits a dynamical feedback mechanism, namely, the $\CP^1$ field determines the effective medium through which the gauge field propagates.

\section{Dynamical Generation of magnetic permeability in the $\CP^1$ theory}

\begin{figure}[ht]
\centering
\begin{tikzpicture}
\begin{feynman}
  \vertex (a) at (-3,0);
  \vertex (b) at ( 3,0);

  \vertex (c) at (-1.2,0);
  \vertex (d) at ( 1.2,0);

  \diagram*{(a) -- [photon, momentum=\(A_\mu(p)\)] (c),
    (d) -- [photon, momentum=\(A_\nu(-p)\)] (b),
    (c) -- [fermion, half left, momentum=\(k\)] (d),
    (d) -- [fermion, half left, momentum=\(k+p\)] (c),};
\end{feynman}
\end{tikzpicture}
\caption{One-loop fermionic vacuum polarization contributing to the gauge-field two-point function.}
\label{fig1}
\end{figure}
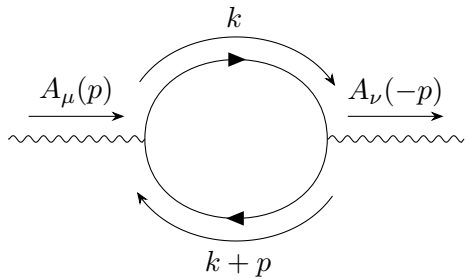

Seeking to generate a contribution of a field-dependent magnetic permeability function in the gauge sector (i.e., the generalization $G(\vert u\vert) F_{\mu\nu} F^{\mu\nu}$), we will adopt a mechanism of dynamical generation. Inspired by the scheme proposed in Refs. \cite{Petrov}, let us consider a Dirac field minimally coupled to the gauge field and endowed with an effective mass that depends on the $\CP^1$ field. This functional dependence allows quantum fluctuations in the fermionic sector to induce, at the one-loop level, corrections to the Maxwell term, thereby yielding an effective dielectric factor governed by the dynamics of the $\CP^1$ field. Therefore, the total action of the system is extended by the inclusion of the spinorial field, naturally leading to the form presented in Eq. \eqref{Eq7}, which serves as the starting point for the radiative generation of the coupling $G(\vert u\vert) F_{\mu\nu} F^{\mu\nu}$. Taking this into account, let us start with the most general theory possible, i.e.,  the case in which the spinorial field is coupled to both $\CP^1$ and vector fields. By employing the mechanism of dynamical generation of new terms, widely used, for instance, in the generation of the Carroll-Field-Jackiw terms \cite{Jackiw}, one obtains the theory described by Eq. \eqref{Eq7} by choosing the effective action, namely, 
\begin{align}\label{Eq10}
    \mathcal{S}=\int d^4 x\, \bar{\psi}\,[i\gamma^\mu \partial_\mu - e \gamma^\mu A_\mu - m(u)]\,\psi + \mathcal{S}^{\mathrm{\CP^1}}[u]+S^{\mathrm{Maxw}}[A],    
\end{align}
with $m(u)$ the dynamics mass. Furthermore, one highlights that the fermionic field acts as a mediator through which the $\CP^1$ dynamics induces effective modifications in the gauge sector. Since our interest lies in the emergence of a field-dependent magnetic permeability, we restrict the analysis to the one-loop fermionic contributions to the gauge-field two-point function. Issues related to renormalizability will not be addressed, once we focus exclusively on the finite quantum corrections responsible for generating magnetic permeability $G(\vert u\vert )F_{\mu\nu}F^{\mu\nu}$. Furthermore, let us assume that the $\CP^1$ field varies slowly in spacetime, allowing us to neglect derivative contributions involving $\partial_\mu u$. Under these assumptions, the leading quantum correction arises from the vacuum polarization diagram depicted in Fig. \ref{fig1}.

Its contribution is given by
\begin{align}\label{Eq11}
    \Gamma^{(2)}=-\frac{e^2}{2}\,\int\,\frac{d^4 p}{(2\pi)^4}\,A_\mu(-p) \Pi^{\mu\nu}(p) A_\nu(p).    
\end{align}

At the one-loop level, the effective quadratic action for the gauge field is governed by the fermionic vacuum polarization tensor. Naturally, this object encodes the response of the fermionic sector to external gauge fluctuations and arises from integrating out the Dirac field in the presence of a background gauge configuration. Diagrammatically, this contribution corresponds to the standard fermionic bubble with two external gauge legs, as shown in Fig. \ref{fig1}. Since the fermions are minimally coupled to the gauge field and possess an effective mass $m(\vert u\vert)$ controlled by the $\CP^1$ background, the resulting polarization tensor depends parametrically on the local value of the scalar field. Under the assumption that the $\CP^1$ field varies slowly in spacetime, derivative corrections involving $\partial_\mu u$ can be consistently neglected, and the fermion mass may be treated as a spacetime-independent parameter within the loop integral. In this regime, the dominant contribution to the effective action is fully captured by the two-point function of the gauge field, whose momentum-space structure is determined by the fermionic  propagators with mass $m(\vert u\vert$). This leads to the standard expression for the vacuum polarization tensor, viz.,
\begin{align}\label{Eq12}
    \Pi^{\mu\nu}(p)=\int \frac{d^d k}{(2\pi)^d}\frac{\mathrm{tr}[\gamma^\mu(\slashed{k}-m(u))\gamma^\nu(\slashed{k}+\slashed{p}-m(u))]}{(k^2-m^2(u))[(k+p)^2-m^2(u)]}.  
\end{align}

To analyze the polarization tensor, let us expand the trace in the numerator. Using the standard identities of the four-dimensional Dirac algebra, i.e., $\mathrm{tr}(\gamma^\mu\gamma^\alpha\gamma^\nu\gamma^\beta)
=d\left(\eta^{\mu\alpha}\eta^{\nu\beta}
-\eta^{\mu\nu}\eta^{\alpha\beta}
+\eta^{\mu\beta}\eta^{\alpha\nu}\right)$ and $\mathrm{tr}(\gamma^\mu\gamma^\nu)=d\,\eta^{\mu\nu}$, from which one obtains
\begin{align}\label{Eq13}
\mathrm{tr}\!\left[\gamma^\mu(\slashed{k}-m)\gamma^\nu(\slashed{k}+\slashed{p}-m)\right]
=d\Big[k^\mu(k+p)^\nu+k^\nu(k+p)^\mu-\eta^{\mu\nu}\big(k\!\cdot\!(k+p)-m^2\big)\Big],
\end{align}
where $m\simeq m(\vert u\vert)$.

After some algebraic manipulations, the denominators take the form $\left[(k+xp)^2-\Delta\right]^2$ with $\Delta=m^2-x(1-x)p^2$. Subsequently, by performing the momentum shift $q=k+xp$, the linear terms in $q$ in the numerator become zero after integration.

Therefore, the polarization tensor boil down to
\begin{align}\label{Eq14}
\Pi^{\mu\nu}(p)=d\int_0^1 dx \int\frac{d^d q}{(2\pi)^d}
\frac{2q^\mu q^\nu-\eta^{\mu\nu}q^2+2x(1-x)\big(p^\mu p^\nu-\eta^{\mu\nu}p^2\big)+\eta^{\mu\nu}m^2}{(q^2-\Delta)^2}.
\end{align}

Naturally, the integral over $q^\mu q^\nu$ can be evaluated by symmetry, leading to
\begin{align}\label{Eq15}
\int d^dq\,\frac{q^\mu q^\nu}{(q^2-\Delta)^2}=\frac{\eta^{\mu\nu}}{d}
\int d^dq\,\frac{q^2}{(q^2-\Delta)^2},
\end{align}
which explicitly guarantees the transversality of the polarization tensor, $p_\mu\Pi^{\mu\nu}(p)=0$.

Using dimensional regularization in $d=4-\epsilon$, the integrals take the form, viz.,
\begin{align}\label{Eq16}
\int\frac{d^dq}{(2\pi)^d}\frac{1}{(q^2-\Delta)^2}
=\frac{i}{(4\pi)^{d/2}}
\Gamma\!\left(2-\frac{d}{2}\right)\Delta^{\frac{d}{2}-2},
\end{align}
and
\begin{align}\label{Eq17}
\int\frac{d^dq}{(2\pi)^d}\frac{q^2}{(q^2-\Delta)^2}
=\frac{i}{(4\pi)^{d/2}}
\frac{d}{2}\,
\Gamma\!\left(1-\frac{d}{2}\right)\Delta^{\frac{d}{2}-1}.
\end{align}

Finally, separating the divergent and finite contributions, one finds that the divergent part is proportional to $(p^\mu p^\nu-\eta^{\mu\nu}p^2)$ and is independent of the field $u$, and can therefore be absorbed into a renormalization of the gauge field. The finite part, on the other hand, depends explicitly on the effective mass $m(\vert u\vert)$ and yields
\begin{align}\label{Eq18}
\Pi^{\mu\nu}_{\text{fin}}(p)=\frac{e^2}{12\pi^2}\big(p^\mu p^\nu-\eta^{\mu\nu}p^2\big)\ln\!\left(\frac{m^2(\vert u\vert)}{\mu^2}\right).
\end{align}

Upon returning to coordinate space, this contribution generates the effective term\footnote{We highlighted that the dielectric permeability function arises from the universal structure of the renormalized fermionic determinant. Thus, the permeability is not merely a function engineered to generate solutions, but rather an effective response of the fermionic vacuum to the $\mathbb{C}\mathrm{P}^1$ background. Moreover, as we shall see below, the scalar sector comes to carry the imprint of fermionic vacuum polarization. The BPS structure does not merely accommodate logarithmic permeability; rather, it encodes this dependence in the BPS decomposition itself.}
\begin{align}\label{Eq19}
\Gamma_{\text{eff}}^{(1)}=-\frac{e^2}{12\pi^2}\int d^4x\,F_{\mu\nu}F^{\mu\nu}\ln\!\left(\frac{m^2(|u|)}{\mu^2}\right),
\end{align}
which unambiguously identifies the emergence of a $\CP^1$-field-dependent magnetic permeability induced by quantum fluctuations.

Thus, the local effective permeability is quantitatively reliable for the topological  vortices, or for a sufficiently heavy fermionic mode. If the effective fermion mass becomes small or
vanishes in the vortex core, or if the core radius becomes comparable to $m^{-1}(u)$,
higher-derivative corrections are no longer negligible and the local expression must be
replaced by the full nonlocal one-loop effective action. After dimensional reduction,
the $(2+1)$-dimensional description is restricted to the zero-mode sector along the
reduced direction and to energies below the corresponding compactification scale.

Naturally, the approximation leading to Eq. \eqref{Eq19} should be understood as the leading term in a local derivative expansion of the fermionic determinant. Its validity requires that the characteristic momenta carried by the $\mathbb{C}\mathrm{P}^1$ background and by the gauge field are small compared with the local fermion gap $m(|u|)$. For the radial BPS configurations, this amounts to requiring $m(f)R_\mathrm{vortex}\gg 1$, or equivalently
\begin{align}
    \max_r\left\vert\frac{m'_f(f)f'(r)}{m^2(f)}\right\vert\ll 1 \qquad \mathrm{and} \qquad \max_r\left\vert\frac{m''_f(f')^2+m'_f(f)f'')}{m^3(f(r))}\right\vert\ll 1.
\end{align}
Thus, in the asymptotic region, where the profiles decay with the scale $m_\mathrm{eff}=-1$, this reduces to $m_\mathrm{eff}/m(\eta_\infty)\ll 1$. Therefore, the effective permeability used in Eq. \eqref{Eq19} is quantitatively reliable for thick vortices, or equivalently for a sufficiently heavy fermionic mode. If $m(f)$ becomes small or vanishes inside the vortex core, or if the core size becomes comparable to $m^{-1}(f)$, derivative corrections are not negligible and the local expression \eqref{Eq19} must be replaced by the full nonlocal one-loop effective action.

\subsection{Dimensional reduction of the effective Maxwell sector}

Having established that quantum fluctuations in the fermionic sector generate, at the one-loop level, a non-polynomial magnetic permeability dependent on the $\CP^1$ field, it is natural to investigate how this effective structure persists in lower-dimensional frameworks. In particular, our analysis focuses on topological configurations in $(2+1)$ dimensions. It is essential to clarify how the effective term derived in $(3+1)$ dimensions will be projected onto $(2+1)$ dimensions. Toward this purpose, let us start by considering the effective contribution induced by quantum corrections in $(3+1)$ dimensions, namely,
\begin{align}\label{Eq20}
    \Gamma_{\text{eff}}^{(1)}=-\frac{e^2}{12\pi^2}\int d^4x\,F_{\mu\nu}F^{\mu\nu}\ln\!\left(\frac{m^2(|u|)}{\mu^2}\right).
\end{align}
The logarithmic dependence of the magnetic permeability is intrinsically tied to the four-dimensional framework.

To study the theory in $3+1$ dimensions, we perform a dimensional reduction by assuming that all fields are independent of the spatial coordinate $x^3$, which has a finite extension $L$. The gauge field is decomposed as $A_\mu = (A_i, A_3)$, with $i = 0,1,2$. Within this framework, the contraction of the electromagnetic tensor splits as
$F_{\mu\nu}F^{\mu\nu} = F_{ij}F^{ij} + 2F_{i3}F^{i3}$, where, under the assumption $\partial_3 A_\mu = 0$, one has $F_{i3} = \partial_i A_3$. Integrating the effective Maxwell action \eqref{Eq20} over the coordinate $x^3$, one obtains 
\begin{align}\label{Eq22}
    \Gamma_{\text{eff}}^{(2+1)}=-\frac{L}{4}\int d^3x\;G(\vert u\vert)\,F_{ij}F^{ij}-\frac{L}{2}\int d^3x\;G(\vert u\vert)\,\partial_i A_3\,\partial^i A_3
\end{align}
with $G(\vert u\vert)=\frac{e^2}{3\pi^2}\mathrm{ln}\left(\frac{m^2(\vert u\vert)}{\mu^2}\right)$.

In the regime relevant, i.e., in $(2+1)$ dimensions, the scalar mode $A_3$ can be consistently truncated (or fixed by an appropriate gauge choice, i.e., $A_3 = 0$), so that the effective Maxwell term in $(2+1)$ dimensions takes its final form 
\begin{align}\label{Eq23}
    \Gamma_{\text{eff}}^{(2+1)}=-\frac{e^2}{12\pi^2 }\int d^3x\;\mathrm{ln}\left(\frac{m(\vert u\vert)^2}{\mu^2}\right)F_{ij}F^{ij}
\end{align}
whose parameter $L$ is unity. Thus, the logarithmic functional profile of the magnetic permeability generated at the quantum level in $(3+1)$ dimensions, after dimensional reduction, is preserved.

\section{Construction in the generalized Maxwell theory}

The Bogomol'nyi, Prasad, and Sommerfield (BPS) approach establishes a precise variational criterion for identifying minimal and localized solutions in gauge theories coupled to scalar fields \cite{Bogomolnyi,Prasad}. By decomposing the energy functional into a sum of positive-definite quadratic terms plus a boundary term, typically associated with a nontrivial homotopy class, one obtains an inequality with the profile $\mathrm{E}\geq\mathrm{E}_{\mathrm{BPS}}$ or $\mathrm{E}\geq\vert Q\vert_{\mathrm{top}}$ (where $\mathrm{E}_{\mathrm{BPS}}$ is the BPS energy, $\vert Q\vert_{\mathrm{top}}$ is the topological charge), whose saturation reduces the second-order Euler-Lagrange equations to a first-order self-dual expressions \cite{Adam,RevAtmaja,RevRamadhan}.
When the bound is saturated, the space of BPS solutions acquires a well-defined geometric structure, described by a finite-dimensional moduli space governed by residual symmetries \cite{RevAtmaja,RevRamadhan,Atmaja,Tong,Gibbons,Lee}. Moreover, in supersymmetric theories, the self-dual configurations are shortened representations of the supersymmetry algebra that saturate the central charge, thereby ensuring nonperturbative protection against renormalization effects \cite{Witten}. Therefore, the BPS approach is a structural mechanism that connects topological stability, the geometry of field space, and nonperturbative dynamics \cite{Witten,Intriligator}.

In light of these considerations, let us analyze the BPS properties from the $\CP^1$-Maxwell theory. Therefore, one adopts the generalized $\CP^1$-Maxwell model, in which the energy is
\begin{align}\label{Eq24}
    \mathrm{E}=\,\int\,d^2x\,\left[\Omega\vert D_iu\vert^2+\frac{G(\vert u \vert)}{2}B^2+\mathcal{\tilde{V}}\right].
\end{align}
where $B$ is the magnetic field and $\Omega$ describes the contribution of the Fubini-Study metric \cite{Manton}. Furthermore, one produces $G(|u|)$ by fermionic vacuum polarization at one loop. In this case, the BPS structure is constructed through a compatible choice of  potential. Thus, these entity should be presented as a BPS structure of the local effective model, rather than as an exact quantum property. Indeed,  $G(|u|)$ arises from one-loop corrections and that the BPS decomposition imposes a rigid relation among $\mathcal{\tilde{V}}$ and $G(|u|)$.

Mathematically, $\Omega$ takes the form
\begin{align}\label{Eq25}
    \Omega=\frac{2}{(1+\vert u\vert^2)^2}.
\end{align}

To examine the BPS property, we reformulated the energy \eqref{Eq24} as
\begin{align}\nonumber
    \mathrm{E}=&\int\,d^2x\,\bigg[\Omega\vert D_1u\pm i D_2u\vert^2+\frac{G}{2}\left(B\mp\frac{\mathcal{W}}{\sqrt{G}}\right)^2-\frac{\mathcal{W}^2}{2}+\mathcal{\tilde{V}}\pm B\mathcal{W}\sqrt{G}\mp i\Omega\varepsilon_{ij}\partial_iu^*\partial_ju+ \\
    \pm&\Omega\partial_i(\varepsilon_{ij}A_j\vert u\vert^2)\mp\frac{2\vert u\vert^2 B}{(1+\vert u \vert^2)^2}\bigg],
    \label{Eq26}
\end{align}
which allows us to conclude that, in order for the systems to exhibit BPS properties, one must require that
\begin{align}\label{Eq27}
    \mathcal{\tilde{V}}=\frac{\mathcal{W}^2}{2} \qquad \mathrm{with} \qquad \mathcal{W}=\frac{2\vert u\vert^2}{\sqrt{G}(1+\vert u\vert^2)^2}.
\end{align}
It is important to emphasize that the relation between the potential $\tilde{\mathcal{V}}$, the permeability function $G(|u|)$, and the auxiliary function $\mathcal{W}(|u|)$ is not an arbitrary assumption. Rather, it follows from the requirement that the static energy \eqref{Eq24} (and Eq. \eqref{Eq26}) admits a Bogomol'nyi decomposition.
From this perspective, $G(|u|)$ is regarded as an input function, fixed by the effective gauge-sector dynamics, whereas $\mathcal{W}(|u|)$ is introduced as an auxiliary function that allows the magnetic contribution to be written as
a perfect square.

Within this framework, the energy functional \eqref{Eq27} boils down to
\begin{align}\label{1Eq27}
    \mathrm{E}=&\int\,d^2x\,\left[\Omega\vert D_1u\pm iD_2u\vert^2+\frac{G}{2}\left(B\mp\frac{\mathcal{W}}{\sqrt{G}}\right)^2\right]+\mathrm{E}_{\mathrm{BPS}},
\end{align}
where
\begin{align}
    \label{Eq28} \mathrm{E}_{\mathrm{BPS}}=&\mp\int\,d^2x\,\left[i\Omega\varepsilon_{ij}\partial_iu^*\partial_ju-\Omega\,\partial_i(\varepsilon_{ij}A_j\vert u \vert^2)\right].
\end{align}

Taking into account the geometric identity of the $\CP^1$ model, viz.,
\begin{align}
    \label{Eq29} \Omega\varepsilon_{ij}\partial_iu\partial_ju=\partial_i\left(\frac{2\varepsilon_{ij}u^*du}{(1+\vert u\vert^2)}\right),
\end{align}
since $\frac{2i\,du*\wedge du}{(1+|u|^2)^2}$ corresponds to the area form on $S^2$, which is closed and locally exact in stereographic coordinates, one concludes that
\begin{align}\label{Eq30}
    \Omega\upepsilon_{ij}\partial_iu^*\partial_ju=\partial_iK_i \qquad \mathrm{with} \qquad K_i=\frac{2\upepsilon_{ij}u^*\partial_ju}{1+\vert u\vert^2}, 
\end{align}
which leads us to
\begin{align}\label{Eq31}
    \mathrm{E}_{\mathrm{BPS}}=\pm\int\,d^2x\, \partial_i(\Omega\upepsilon_{ij}A_j\vert u\vert^2-K_i).
\end{align}

 Thus, when the energy saturates the BPS bound, $\mathrm{E}=\mathrm{E}_{\mathrm{BPS}}$, one finds
\begin{align}\label{Eq32}
    D_1u=\mp iD_2u \qquad \mathrm{and} \qquad
    B=\pm\frac{2\vert u\vert^2}{G(\vert u\vert)(1+\vert u\vert^2)^2}.
\end{align}
Note that the energy [Eq. \eqref{Eq24}] of the generalized $\mathrm{CP}^1$–Maxwell model is structured by three fundamental dynamical contributions, i.e., the nonlinear kinetic term $\Omega\, \vert D_i u\vert^2$ in target space, which encodes the spherical target geometry through a $\vert u\vert$-dependent metric factor; the generalized gauge term $\frac{G(\vert u\vert)}{2} B^2$, which introduces an effective magnetic permeability depending on the $\CP^1$ field; and the effective potential $\tilde{V}$, whose form must be rigidly selected to allow for a consistent of the BPS property. The presence of the generalized functions $\Omega(\vert u\vert)$ and $G(\vert u\vert)$ implies a nontrivial modification of the standard energy structure of the Maxwell-Higgs model \cite{Bazeia1,Lima1}, simultaneously altering the metric on configuration space and the dynamics of the gauge sector. In particular, the possibility of rewriting the functional as a sum of perfect squares plus a boundary term imposes a specific functional relation among $\tilde{\mathcal{V}}$, $G$, and the superpotential function $\mathcal{W}$, ensuring that the saturation limit reduces the equations of motion to first-order self-dual expressions \cite{Bogomolnyi,Prasad}.

\subsection{Radially symmetric BPS configurations}

To proceed with the analysis of self-dual solutions in the generalized $\CP^1$-Maxwell model, it is necessary to explicitly investigate the field configurations that saturate the BPS bound and possess nontrivial topological character. Particularly, we are interested in finite-energy, spatially localized solutions that are invariant under rotations in the plane and that may, a priori, describe structures resembling the vortices. Such configurations are naturally characterized by axial symmetry, allowing the angular dependence of the fields due to a phase associated with the winding number  \cite{Rajaraman,Manton2004}. In this framework, we adopted a Nielsen-Olesen-like radial ansatz for the $\CP^1$ scalar field \cite{Nielsen}, together with a radially symmetric gauge-field ansatz introduced in Refs. \cite{VachaspatiB,Jackiw00}, which leads us to expect to capture topological configurations for the generalized $\CP^1$-Maxwell generalized by the magnetic permeability function. In pursuit of this purpose, we will adopt the ans\"atze
\begin{align}\label{Eq33}
    u=f(r)\,\mathrm{e}^{in\theta} \hspace{0.5cm} \mathrm{and} \hspace{0.5cm} A_\theta=\frac{[n-a(r)]}{\mathrm{e}\,r}.
\end{align}
Here, $f(r)$ and $a(r)$ denote the radial profile functions associated with the $\CP^1$ and the gauge fields, respectively. Moreover, $n \in \mathbb{Z}$ is the winding number, and $\theta$ is the angular coordinate.

Naturally, the ansatz \eqref{Eq33} implies that the corresponding field configurations support a nonvanishing magnetic field, $B = F_{12}$ \cite{Jackiw}. That, in turn, allows us to conclude that the associated magnetic flux ($\Phi_{\mathrm{flux}}$) is
\begin{align}
    \label{Eq34}
    \Phi_{\mathrm{flux}}=\frac{2\pi}{e}[a(0)-a(\infty)].
\end{align}
In this framework, it becomes essential to examine the Gauss law. Accordingly, by considering the generalized theory defined in \eqref{Eq23}, we inspect the equation of motion \eqref{Eq9}, which follows from the variation concerning the gauge field. By analyzing the zeroth component of Eq. \eqref{Eq9}, one obtains the Gauss law, i.e., 
\begin{align}\label{Eq35}
    J^0=\partial_\mu[G(\vert u\vert)F^{\mu\, 0}]=\frac{2ie}{(1+\vert u\vert^2)^2}[u^*\,(D^0u)-(D^0u)^*u]=0.
\end{align}
Therefore, in the static regime, one has $A_0 = 0$, which implies that only magnetic structures can arise within the theory.

To proceed, let us impose the topological boundary conditions \cite{Cunha}, viz.,
\begin{align}\label{Eq36}
    f(0)=\eta_0, \quad f(\infty)=\eta_\infty, \quad a(0)=\beta_0, \quad a(\infty)=\beta_\infty. 
\end{align}
In this case, $n_0$ and $n_\infty$ are integers, while $\eta_{0,\infty} \in \mathbb{R}$. Under these boundary conditions, the magnetic flux of the configurations boils down to
\begin{align}\label{Eq37}
    \Phi_{\text{flux}}=\frac{2\pi M}{e}
\end{align}
with $M=\beta_0-\beta_\infty>0$ and $M\in \mathbb{Z}_+$.

Furthermore, by assuming the ansatze \eqref{Eq33}, one obtains the BPS energy density, i.e.,
\begin{align}
    \label{Eq37.1}
    \mathcal{E}_{\mathrm{BPS}}=\frac{4f'^2}{G(f)(1+f^2)^4}.
\end{align}

The adoption of a Nielsen-Olesen-like ansatz \cite{Nielsen} for the $\CP^1$ field, together with the ansatz for the gauge sector, allows the first-order equations \eqref{Eq32} be reduced to self-dual equations, viz.,
\begin{align}\label{Eq38}
    f'=\mp\frac{af}{r}\qquad \mathrm{and} \qquad \frac{a'}{r}=\pm\frac{2f^4}{G(\vert f\vert)(1+f^2)^2}
\end{align}
 Naturally, the BPS approach preserves the topological quantization of the magnetic flux, $\Phi_{\mathrm{flux}}=2\pi M/\mathrm{e}$. Meanwhile, the Gauss law enforces $A_0=0$ in the static regime, restricting the spectrum of solutions to purely magnetic configurations whose stability is guaranteed. Physically, the magnetic permeability term $G(|f|)$ modifies the effective coupling of the magnetic field to the core, affecting both the width of the magnetized structure and the asymptotic profile of the solutions. Consequently, the interplay between the Fubini-Study metric and the magnetic permeability gives rise to a more general class of topological defects, in which the internal structure is sensitive to the modifications of the $\CP^1$ field. One highlights that the case without magnetic permeability has been recently reported in Ref. \cite{Cunha}.

 \section{On the magnetized configurations}

 \subsection{General asymptotic behavior}

 \subsubsection{The case near the origin}

Let us now examine the asymptotic regimes of the solutions. To ensure the regularity of the $\CP^1$ field in $r \to 0$, one must require a regular profile at the origin. This requirement entails that the modulus of $f(r)$ vanishes sufficiently fast for $n \neq 0$. Furthermore, since the topological boundary conditions enforce $a(0) = \beta_0$, it follows that
 \begin{align}\label{Eq40}
    f\simeq \mathcal{C}_0\,r^{\vert \beta_0\vert}.
 \end{align}
Therefore, the profile of the $\CP^1$ sector vanishes as a power law determined by the value of the gauge field at the origin, exactly as in the case of Nielsen-Olesen vortices \cite{Nielsen}.

We now analyze in greater detail the profile $a(r)$ when $r\to 0$. Regularity of the solution at the origin requires the gauge field to remain finite and sufficiently smooth in this regime. Taking this into account, one obtains
\begin{align}\label{Eq41.1}
    \frac{a'}{r}\sim\frac{2f^2}{G(f)(1+f^2)^2}.
\end{align}
Since $f\to 0$ as $r\to 0$ (see Eq.~\eqref{Eq40}), one finds that $G(f)(1+f^2)^2\simeq G\vert_{f=0}=G(0)$. It therefore follows that
\begin{align}\label{Eq41}
    a(r)\simeq n_0+\mathcal{O}(r^{2\vert \beta_0\vert+2}).
\end{align}
Therefore, near the origin, $f(r) \sim r^{|n_0|}$ and $a(r) \sim n_0$ in a smooth manner, is precisely what ensures the topological and energetic regularity of the solution. The factor $r^{|\beta_0|}$ guarantees that the scalar field vanishes sufficiently rapidly as $r \to 0$, thereby preventing divergences in the angular kinetic term, which contains contributions proportional to $(\eta_0-a)^2f^2/r^2$. Simultaneously, the fact that $a(r)$ approaches $\eta_0$ smoothly implies that $a'(r) \to 0$ at the core, so that the magnetic field $B(r) \sim -a'/r$ remains finite. This structure is fully consistent with the boundary conditions dictated by field regularity. Consequently, the core of the configurations is free from physical singularities.

 \subsubsection{The behavior at spatial infinity}

Meanwhile, in the asymptotic regime $r \to \infty$, the boundary condition displayed in Eq.~\eqref{Eq36} requires that, far from the core structure, the fields approach their vacuum values, viz., $f(\infty)=\eta_\infty$ and $a(\infty)=\beta_\infty$. Accordingly, at spatial infinity, one may parametrize the scalar profile as $f(r)=\eta_\infty-\delta f(r)$, with $\delta f(r)\ll 1$, which allows us to write\footnote{One highlights that the BPS vortex size must satisfy $m(f)R_{\rm vortex}\gg 1$. In the asymptotic region, where the profiles decay with the characteristic scale $m_{\rm eff}^{-1}$, this condition reduces to $m_\mathrm{\mathrm{eff}}\ll m(\eta_\infty)$.}
 \begin{align}\label{Eq42}
     f'\simeq\mp\frac{\eta_\infty\beta_\infty}{r},
 \end{align}
Consequently, one obtains the asymptotic behavior, $\delta f(r)\sim \frac{\eta_\infty \beta_\infty}{r}$, which enables us to consistently approximate the scalar profile in the large-$r$ regime and proceed with the linearized analysis of the corresponding field equations, viz.,
\begin{align}\label{Eq43}
    f(r)\simeq \eta_\infty-\frac{\eta_\infty\beta_\infty}{r}+\dots \qquad \mathrm{with} \qquad r\to \infty.
\end{align}

Simultaneously,
\begin{align}\label{Eq44}
    \frac{a'}{r}\to\frac{2\beta_\infty^2}{G(\eta_\infty)(1+\beta_\infty^2)^2}\equiv M^2,
\end{align}
where $M^2$ is a constant parameter. Furthermore, upon examining the effects induced by the magnetic permeability function, one can consistently assess how the generalized gauge-sector dynamics modify the asymptotic structure of the solutions and, in particular, the effective mass scale governing the linearized fluctuations.
\begin{align}\label{Eq45}
    G(\eta_\infty)=\frac{e^2}{3\pi^2}\mathrm{ln}\left(\frac{m^2(\eta_\infty)}{\mu^2}\right)    
\end{align}
One notes that for $m(\eta_\infty)\in(0,\infty)$ the dielectric function $G(\eta_\infty)$ remains finite. In this regime, the field variables exhibit a power-law decay of the form
\begin{align}\label{Eq46}
    \delta f,\;\;\delta a\sim \mathrm{e}^{-Mr}.
\end{align}
Conversely, if $m(\eta_\infty)\to 0$, the asymptotic profile ceases to be exponential and becomes polynomial. Therefore, when the effective mass vanishes in the vacuum configuration, the dielectric function diverges logarithmically, and the magnetic structure acquires a long-range character.

\subsection{On the linear stability}

Hereafter, let us examine the stability of the configurations. To accomplish our purpose, we require a finite energy and magnetic flux profile for stability. Therefore, one must impose $D_i u = 0$ and $B = 0$ at spatial infinity. Within this framework, the consistency conditions reduce to $f'(r)\to 0$ and $a(\infty),\eta_\infty \to 0$. Thereby, two distinct branches then emerge. The first corresponds to the non-topological sector, arising when $\eta_\infty = 0$ while $a(\infty)\neq 0$. In this case, the magnetic flux is non-topological with asymptotic decay of the profiles $f(r)$ and $a(r)$, polynomial omnipresent, being determined by the large-field behavior of the dielectric function $G(\vert f \vert)$. Our primary interest consists of the topological configurations, for which $\eta_\infty \neq 0$ and $a(\infty)=0$. For this case, since $\eta_\infty \neq 0$, we may consistently linearize the BPS equations \eqref{Eq38} around the vacuum, which leads to
\begin{align}\label{Eq47}
    (\delta f)'\simeq\pm\frac{\eta_\infty}{r}\delta a \qquad \mathrm{and} \qquad \frac{\delta a'}{r}\simeq\mp\frac{4\eta_\infty}{G(\eta_\infty^2)(1+\eta_\infty^2)^2}\delta f
\end{align}

Upon decoupling the equations \eqref{Eq47}, one derives
\begin{align}\label{Eq48}
    \delta f''+\frac{1}{r}\delta f'-m_{\mathrm{eff}}^2\,\delta f=0,
\end{align}
where
\begin{align}\label{Eq49.1}
    m_\mathrm{eff}^2=\frac{4\eta_\infty^2}{G(\eta_\infty)(1+\eta_\infty^2)^2}.
\end{align}
Note that Eq. \eqref{Eq48} corresponds to the modified Bessel equation of zeroth-order. Therefore, the physically admissible solution in the asymptotic region is 
\begin{align}
    \label{Eq49}
    \delta f(r)\sim K_0(m_{\mathrm{eff}}\,r),
\end{align}
Hence, for $r \to \infty$, one obtains
\begin{align}\label{Eq51}
    K_0(m_\mathrm{eff}\,r)=\sqrt{\frac{\pi}{2mr}}\,\mathrm{e}^{-mr}.
\end{align}
Consequently,
\begin{align}\label{Eq52}
    \delta f,\;\delta a\sim\mathrm{e}^{-m_{\mathrm{eff}}\,r}.
\end{align}
Therefore, in the asymptotic regime, the linearization of the BPS equations around the vacuum configuration defined by $f(\infty)=\eta_\infty\neq 0$ and $a(\infty)=0$ leads to a modified Bessel-like differential equation for the scalar fluctuation. The physically admissible solution is proportional to $K_0(m_{\mathrm{eff}}, r)$. The parameter $m_{\mathrm{eff}}$ naturally emerges as the mass scale associated with the coupled $CP^1$–gauge mode. Consequently, the asymptotic profiles decay exponentially, $\delta f,\; \delta a \sim \mathrm{e}^{-m_{\mathrm{eff}} r}$, thereby ensuring magnetic flux confinement and finite energy. This result demonstrates that the magnetic permeability function plays a direct role in determining the characteristic length scale of the solution, viz., $l \sim 1/m_{\mathrm{eff}}$.

Meanwhile, one notes that larger values of $G(\eta_\infty)$ reduce the effective mass and consequently generate broader vortex-like structures, whereas smaller values compress the magnetic core. Accordingly, the generalization of the Maxwell sector through the function $G(f)$ not only preserves the topological character of the BPS solutions but also introduces a dynamical mechanism that controls the defect width. In this form, one establishes an explicit connection between the structure of the effective vacuum and the intrinsic spatial scale of the self-dual configuration.

\section{Solutions of the self-dual equations}

The system of self-dual equations announced in Eq. \eqref{Eq38} completely determines the radial dynamics of the topological configurations in the generalized $\CP^1$-Maxwell model. The structure of these expressions reveals that the profile of the solutions depends sensitively on the functional shape of the magnetic permeability $G(\vert f\vert)$, which controls the effective coupling between the scalar field and the magnetic sector. Particularly, different choices for the mass profile $m(\vert f \vert)$, responsible for the dynamical generation of $G(\vert f\vert)$, can lead to distinct physical regimes, modifying both the width of the magnetic core and the asymptotic structure. To make these properties explicit and to understand the effects of induced permeability on the topological defects, we shall analyze some representative cases.

\subsection{Example 1: The case without permeability, $m(f)=\left\vert\mu\sqrt{\mathrm{e}^{\frac{32\pi^2}{e^2}}}\right\vert$}

For the first example, we consider the simplest case, in which the effective mass of the fermionic field is constant, i.e., $m(f)=\left\vert\mu\sqrt{\mathrm{e}^{\frac{32\pi^2}{e^2}}}\right\vert$. In this case, the quantum correction responsible for generating the magnetic permeability becomes trivial, resulting in a constant effective factor in the Maxwell sector. Therefore, the permeability function reduces to $G = 1$, recovering the na\"{i}ve $\CP^1$-Maxwell model without additional generalizations in the magnetic coupling. Naturally, this scenario provides a relevant reference point for investigations into the self-dual equations, as it reproduces the well-known structure of the BPS equations concerning the $\mathbb{CP}^1$ model. Thus, alongside the effects introduced by field-dependent $\CP^1$ permeabilities, we also examine this limiting case, which allows us to characterize the standard behavior of the solutions and to establish a comparative baseline for the generalized regimes discussed in the subsequent subsections.

Let us begin by considering the case in which the mass of the fermionic coupling assumes the profile   
\begin{align}\label{Eq53}
    m(f)=\left\vert\mu\sqrt{\mathrm{e}^{\frac{32\pi^2}{e^2}}}\right\vert,
\end{align}
i.e., the system of self-dual equations presented in Eq. \eqref{Eq54} describes the regime in which the magnetic permeability becomes constant, $G = 1$, so that the dynamics of the gauge sector recovers the standard form of the usual $\CP^1$–Maxwell model. In this case, the first-order equations reduce to
\begin{align}\label{Eq54}
    f'=\pm\frac{af}{r} \qquad \mathrm{and} \qquad \frac{a'}{r}=\mp\frac{2f^2}{(1+f^2)^2}.
\end{align}
Note that the system of BPS equations was recently reported in Ref. \cite{Cunha}. Moreover, these fields fully encode the structure of the topological configurations that saturate the BPS bound. The first equation establishes a direct coupling between the radial profile of the scalar field $f(r)$ and the gauge function $a(r)$, reflecting the fact that the spatial variation of the $\CP^1$ field is compensated by the gauge connection to minimize the energy of the system. The second equation relates the curvature of the gauge field, i.e., the magnetic field $B\sim -a'/r$, to the effective density of the scalar field in the target space, modulated by the geometric factor $\Omega$, which corresponds to the Fubini-Study metric of the space $\mathbb{S}^2$. 

\begin{figure}[!ht]
    \centering
    \subfigure[Numerical solution of $f(r)$ vs. $r$.]{\includegraphics[width=8cm,height=6cm]{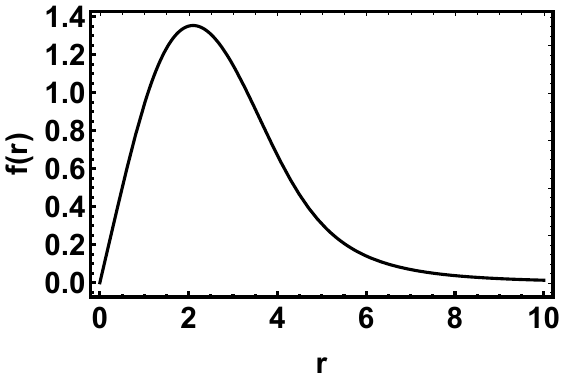}}\hfill
    \subfigure[Numerical solution of $f(r)$ vs. $r$.]{\includegraphics[width=8cm,height=6cm]{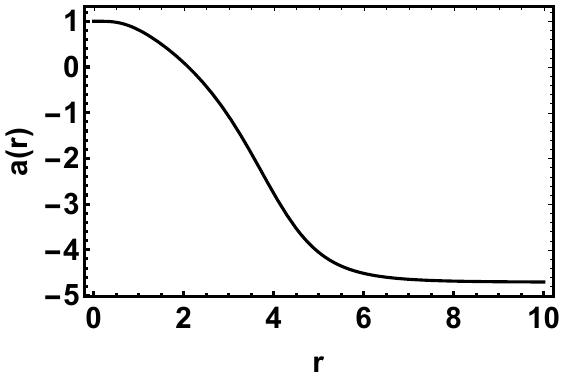}}
    \caption{Numerical solutions of the variable fields with unit winding number.}
    \label{fFig1}
\end{figure}

To make the results more transparent, let us numerically evaluate the BPS equations presented in Eq. \eqref{Eq54}. By performing the numerical integration, we obtain the profiles shown in Fig. \ref{fFig1} for both the $\CP^1$ field and the gauge field.
Figure \ref{fFig1} displays the numerical solutions for the profile functions $f(r)$ and $a(r)$ that solve the self-dual system given in Eq. \eqref{Eq54}, in the regime where the magnetic permeability is constant (i.e., $G=1$). Figure \ref{fFig1}(a) shows the profile of the scalar field $f(r)$, while Fig. \ref{fFig1}(b) presents the profile of the gauge function $a(r)$, both obtained for unit winding number. One notes that the scalar field grows monotonically from the origin, where it satisfies the regularity condition $f(0)=0$, and asymptotically approaches its vacuum value at large distances. This behavior is characteristic of lump-like configurations and reflects the fact that the $\CP^1$ field smoothly interpolates between the defect core and the same vacuum state. The initial growth rate is consistent with the asymptotic analysis near the origin, $f(r)\sim r^{\vert n\vert}$, ensuring the regularity of the kinetic contributions in the limit $r\to 0$.

On the other hand, Fig. \ref{fFig1}(b) shows that the gauge function $a(r)$ starts from an integer value at the origin and decreases monotonically toward zero as $r\to \infty$. This behavior ensures the quantization of the magnetic flux, constituting precisely the mechanism responsible for compensating the angular phase of the scalar field in the ansatz $u=f(r)\mathrm{e}^{in\theta}$. Consequently, the magnetic field remains localized within a finite region of space, characterizing the presence of a magnetized ring-like profile, see Ref. \cite{Cunha}. The monotonic behavior of $a(r)$ further indicates that the magnetic flux is progressively expelled outward from the core as the system approaches the vacuum.

\begin{figure}[!ht]
    \centering
    \subfigure[Numerical solutions: $B(r)$ vs. $r$.]{\includegraphics[width=8cm,height=5.5cm]{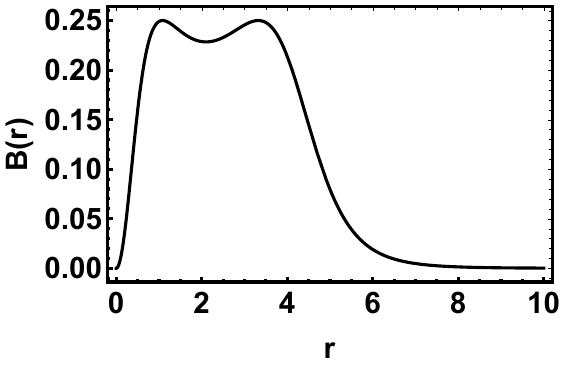}}
    \subfigure[Planar magnetic configuration.]{\includegraphics[width=5.5cm,height=5.5cm]{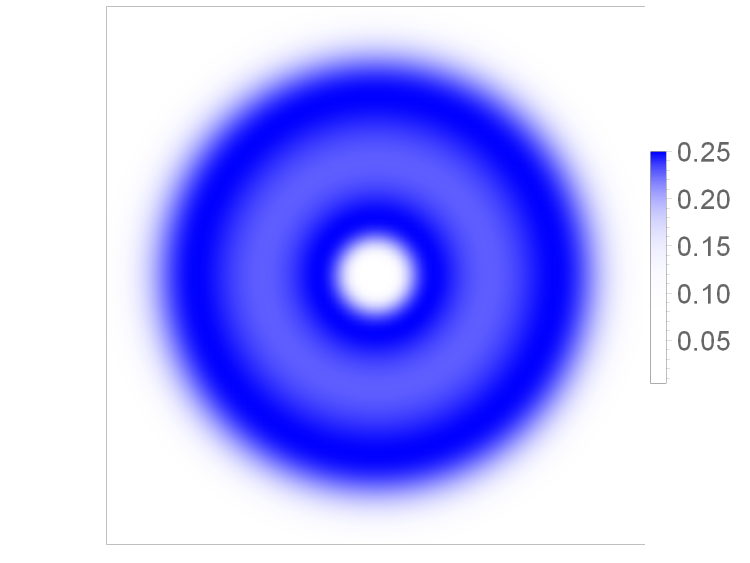}}
    \caption{Magnetic field $B(r)$ vs. $r$ with unit winding number.}
    \label{Fig2}
\end{figure}

Meanwhile, Fig. \ref{Fig2} presents the radial profile of the magnetic field $B(r)$ associated with the self-dual solutions from Eq. \eqref{Eq54} together with the definition of the magnetic field $B=F_{12}$ \cite{Jackiw00}. Since, within the radial ansatz, the magnetic field is $B(r)=-a'(r)/r$, its spatial distribution is directly related to the behavior of the gauge function $a(r)$ discussed previously. The plot displayed in Fig. \ref{Fig2}(a) shows that the magnetic field reaches a maximum value in a region close to the defect core and rapidly decays as $r$ increases, approaching zero in the asymptotic regime. This behavior indicates that the magnetic flux remains confined within a finite region of space, characterizing the presence of a localized structure resembling the magnetized lumps.

Physically, the fact that $B(r)$ exhibits a critical point near the defect core reflects the mechanism of magnetic flux confinement imposed by the BPS equations. Near the origin, the scalar field becomes suppressed due to the regularity condition $f(0)=0$, which allows the magnetic field to attain a larger intensity in this region. As the scalar field increases radially and approaches its vacuum value, the second self-dual equation enforces a gradual decrease in the magnetic intensity, promoting a redistribution of the flux until it vanishes at spatial infinity.

Finally, the planar configuration shown in Fig. \ref{Fig2}(b), corresponding to the magnetic field distribution, allows us to visualize the spatial structure of the solution in the two-dimensional plane. One notes an approximately circularly symmetric distribution, consistent with the radial ansatz adopted for the fields, together with the formation of concentric ring-like structures of magnetic induction. The maximum intensity localized near the center and the smooth radial decay indicate that the defect is magnetized and well defined, whose width is determined by the dynamical scale of the model. Taken together, these results confirm that the $\CP^1$-Maxwell system in the BPS regime supports magnetized lump-like configurations with quantized magnetic flux \cite{Cunha}.

\begin{figure}[!ht]
    \centering
    \subfigure[Plot of $\mathcal{E}_{\mathrm{BPS}}(r)$ vs. $r$]{\includegraphics[width=8cm,height=5.5cm]{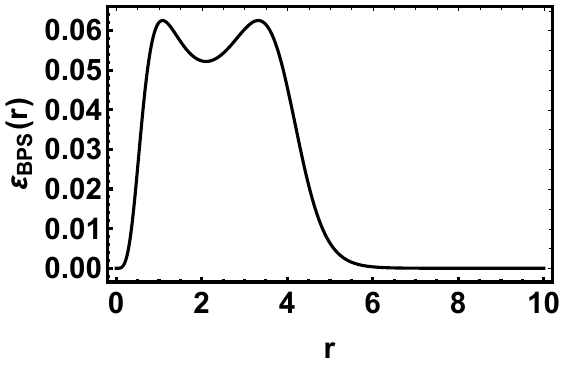}}
    \subfigure[Planar profile for BPS energy density.]{\includegraphics[width=5.5cm,height=5.5cm]{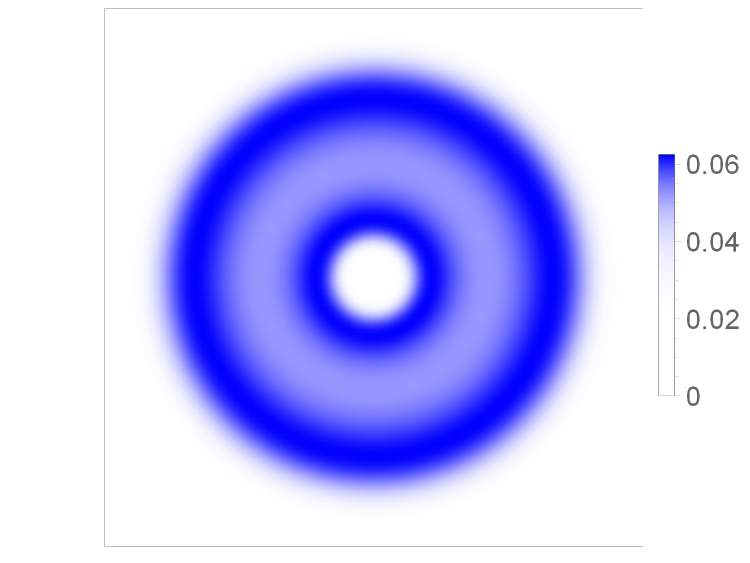}}
    \caption{BPS energy density $ \mathcal{E}_{\mathrm{BPS}}(r)$ vs. $r$ with unit winding number.}
    \label{Fig3}
\end{figure}

By substituting the numerical solutions into the energy expression \eqref{Eq37.1}, one obtains the BPS energy density. Figure \ref{Fig3} displays the behavior of the BPS energy density concerning the self-dual solutions of Eq. \eqref{Eq54}. Figure \ref{Fig3}(a) shows the radial profile of the energy density. Meanwhile, the plots \ref{Fig3}[(b)–(e)] illustrate the corresponding distribution in the plane. The results show that the energy density is strongly concentrated within a finite region of space, reaching a maximum near the vortex core and decaying to zero when $r$ increases.

From a physical viewpoint, the peak of the energy density in the central region is associated with the strong spatial variation of the fields in this domain. Near the origin, the scalar field grows rapidly from $f(0)=0$. Meanwhile, the gauge field undergoes a significant variation to compensate for the angular phase of the $\CP^1$ field. This interplay between the scalar gradient and the magnetic energy generates a highly energized region that defines the core of the ring-like magnetized lumps. As the system approaches the asymptotic regime, both fields gradually converge to their vacuum values. This behavior suppresses the kinetic and magnetic contributions, causing the energy density to vanish at spatial infinity. The ring-like profiles displayed in Fig. \ref{Fig3}(b) provide a planar representation of the energy density, allowing a visualization of the geometry of the solution in the plane.

\subsection{Example 2: The logarithmic case $m(f)=\left\vert\mu\sqrt{P(f)^\frac{32\pi^2}{e^2}}\right\vert$}

Having analyzed the limiting case in which the magnetic permeability is constant, we now turn to the investigation of scenarios where the function $G(\vert f\vert)$ exhibits a nontrivial dependence on the scalar field. Within this regime, the effective interaction between the $\CP^1$ and gauge sectors is modified, producing changes in the structural profile of the self-dual solutions. Particularly, the presence of a field-dependent permeability with logarithmic behavior introduces new dynamical mechanisms capable of controlling the spatial distribution of the magnetic field and, consequently, the width of the vortex core. Different choices for the functional profile of $G(\vert f\vert)$, influenced by the behavior of the effective mass $m(\vert f\vert)$, can lead to distinct physical regimes where the internal properties may be significantly altered.

\begin{figure}[!ht]
    \centering
    \subfigure[Numerical solution of the field variable $f(r)$.]{\includegraphics[width=8cm,height=6cm]{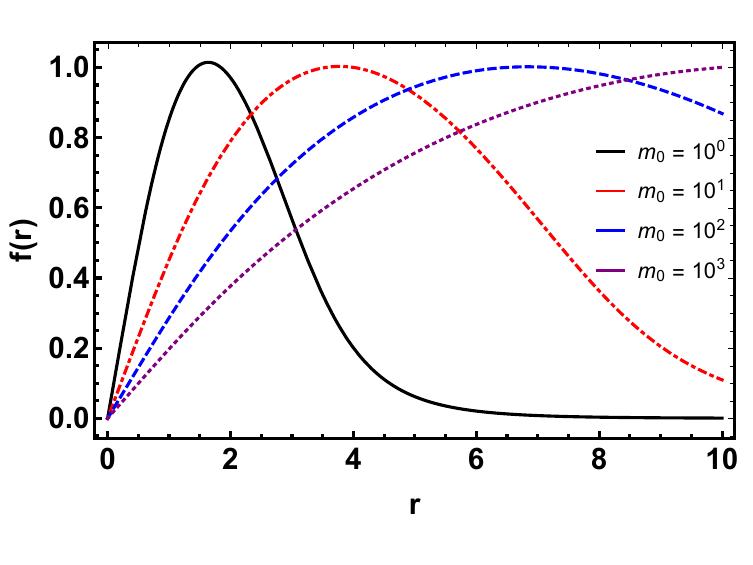}}\hfill
    \subfigure[Numerical solution of the field variable $a(r)$.]{\includegraphics[width=8cm,height=6cm]{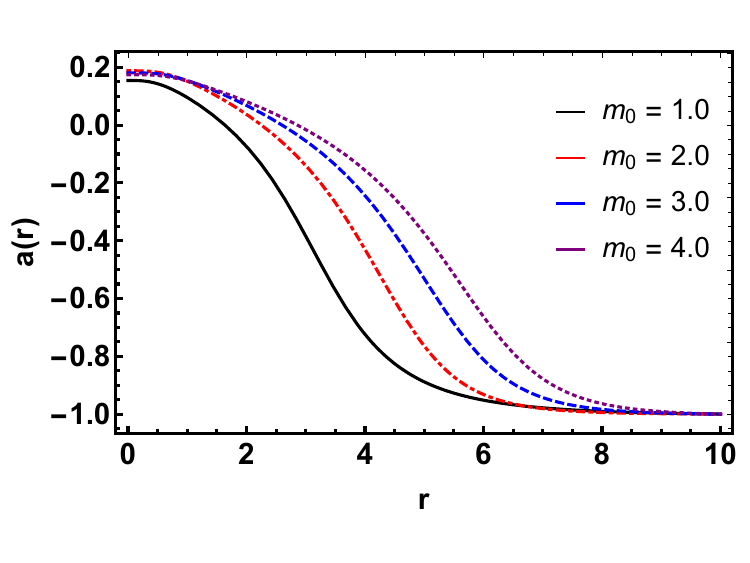}}
    \caption{Numerical solutions with coupling $e=1$ and unit winding number.}
    \label{Fig4}
\end{figure}

To proceed with our analysis, let us consider the case $m=\left\vert\mu\sqrt{P(f)^\frac{32\pi^2}{e^2}}\right\vert$, which leads to a magnetic permeability of the form $G=\mathrm{ln}(\vert P(f)\vert)$. This choice leads us to the BPS equations, viz.,
\begin{align}\label{Eq56}
    f'=\pm\frac{af}{r} \qquad \mathrm{and} \qquad \frac{a'}{r}=\mp\frac{2f^2}{\mathrm{ln}\left(\vert P(f)\vert\right)(1+f^2)^2}
\end{align}
where $P(\vert f\vert)$ is an arbitrary polynomial. For simplicity, let us consider the simplest case, i.e., $P(f)=m_0+f(r)$, where $m_0$ is a positive and constant mass parameter. Taking this into account, let us now examine the solutions of Eq. \eqref{Eq56}. One displays the corresponding numerical solutions in Fig. \ref{Fig4}.

Naturally, one initially notes that the field profiles exhibit the behavior expected from the topological boundary conditions, i.e., the scalar field vanishes smoothly at the origin and approaches the vacuum value at spatial infinity. Meanwhile, the gauge function interpolates monotonically between the integer value, which can be associated with the winding number at the origin, and the vanishing value in the asymptotic regime. This behavior confirms the regularity of the solution throughout the entire spatial domain, ensuring the finiteness of the total energy. Additionally, the effective width of magnetic vortices is influenced by magnetic permeability. In particular, modifications in these terms alter the rate at which the fields relax toward the vacuum, producing either more compact or broader vortices depending on the regime under consideration. Consequently, the region where the energy and the magnetic field concentrate is also modified, directly reflecting changes in the internal structure of the vortex.

Another relevant aspect highlighted by the results in Fig. \ref{Fig4} is that the field profiles remain smooth and monotonic throughout the entire radial domain, a characteristic feature of stable BPS solutions. This behavior reinforces the consistency of the first-order equations, as well as the compatibility between the structure of the effective potential and the conditions required forthe BPS bound. These results confirm that the model supports well-defined topological configurations whose geometry and spatial extension can be controlled by the parameters introduced in the generalized framework.
\begin{figure}[!ht]
    \centering
    \subfigure[Profile of the magnetic field vs. $r$.]{\includegraphics[width=8cm,height=6cm]{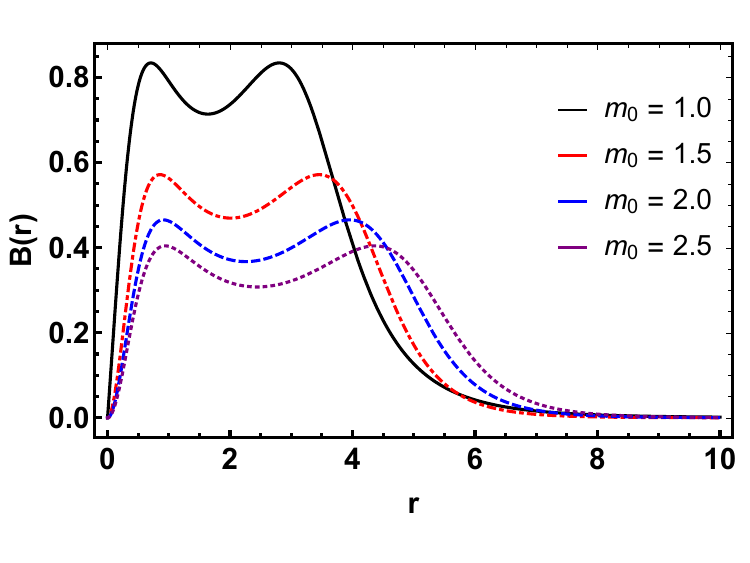}}\\
    \subfigure[The case $m_0=1.0$]{\includegraphics[width=4cm,height=4cm]{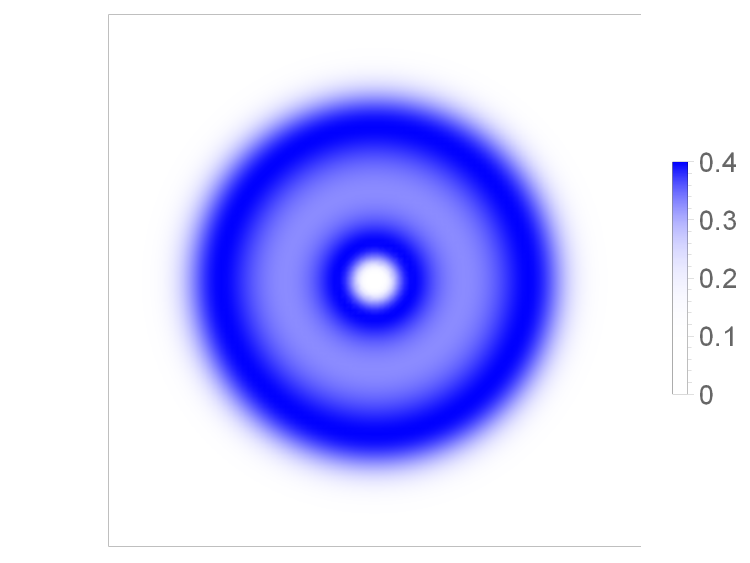}}\hfill
    \subfigure[The case $m_0=1.5$]{\includegraphics[width=4cm,height=4cm]{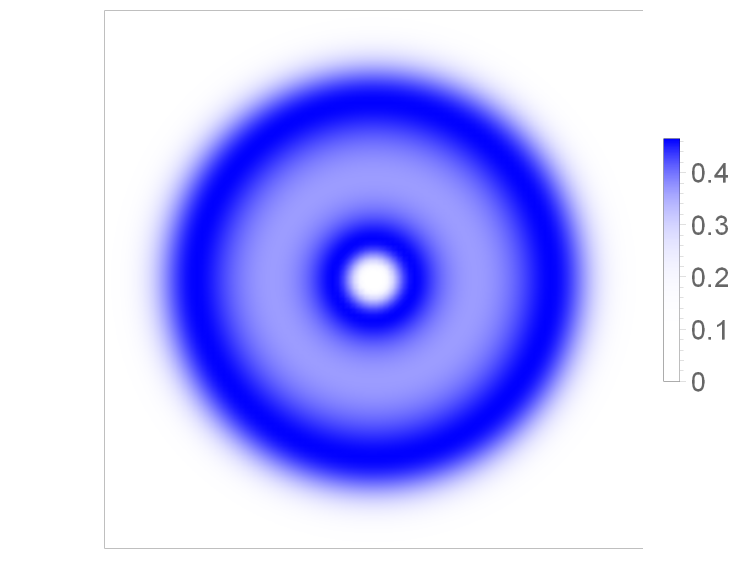}}\hfill
    \subfigure[The case $m_0=2.0$]{\includegraphics[width=4cm,height=4cm]{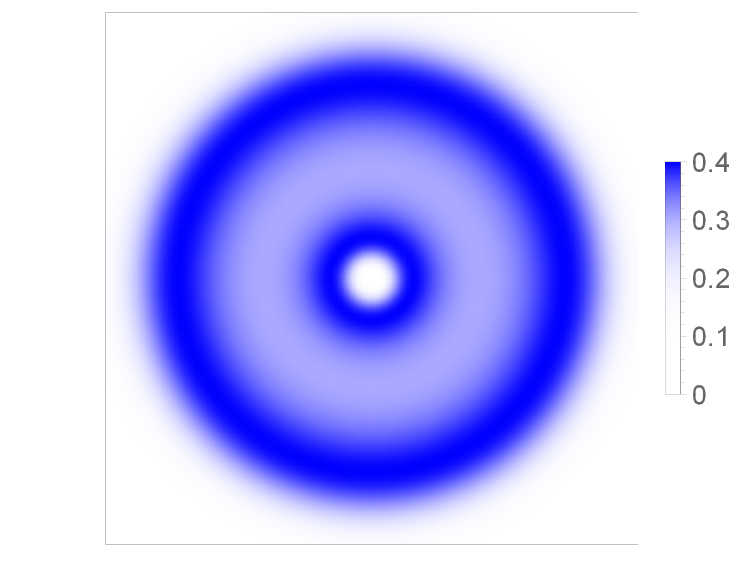}}\hfill
    \subfigure[The case $m_0=2.5$]{\includegraphics[width=4cm,height=4cm]{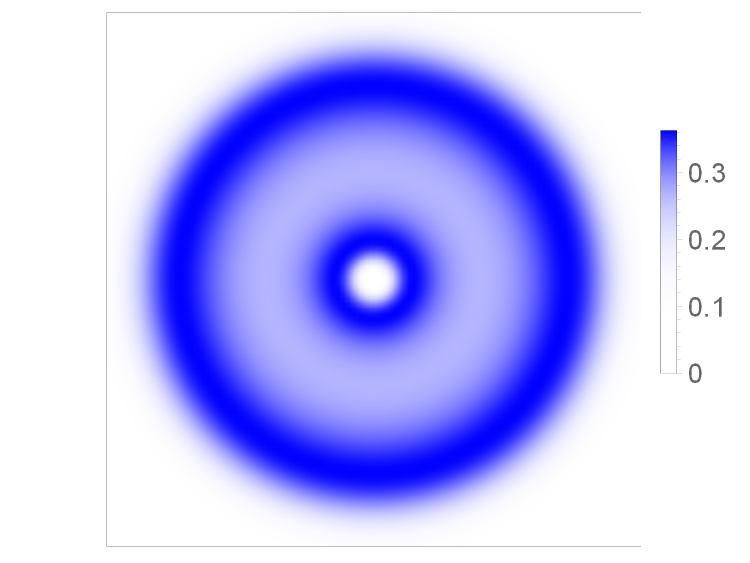}}
    \caption{Numerical solutions for the magnetic field.}
    \label{Fig5}
\end{figure}
Figure \ref{Fig5} displays the radial distribution of the physical quantities associated with the BPS solutions, providing a direct visualization of the internal structure of the vortex configurations. One notes that the magnetic field is concentrated within a finite region of space, characterizing the localized nature of the topological defect. Particularly, the distribution exhibits a well-defined maximum near the vortex core, followed by a smooth decay as $r$ increases, reflecting the relaxation of the solutions toward the vacuum state in the asymptotic regime.
This behavior indicates that the energy and the associated fields remain confined within a limited spatial region, ensuring the finiteness of the total energy. Moreover, the shape of the profile shows that the generalized terms of the model directly influence the effective width of the structure, controlling the geometric expansion of the magnetic vortices and the region where the magnetic induction becomes more intense. In this way, the results displayed in Fig. \ref{Fig6}[(a)–(e)] reinforce the consistency of the obtained solutions and illustrate how the dynamics of the model determine the spatial organization of the vortex configurations in the BPS regime.

Finally, by considering the self-dual equations \eqref{Eq56} together with Eq. \eqref{Eq28}, one obtains the BPS energy density. The corresponding profile of the BPS energy density in the model under consideration allows for a clearer analysis of the energy distribution. One perceives that the profiles remain radially localized, with the dominant contributions concentrated within a finite region around the vortex core and decaying smoothly as $r \to \infty$, where the solutions approach the vacuum state. This behavior reflects the localized nature of the topological configuration, guaranteeing the finiteness of the total energy. Moreover, the profiles indicate that the internal structure of the vortex is sensitive to the parameters controlling the magnetic permeability, which directly influence the effective width of the defect and the location where the physical distributions attain their maximum values. Thereby, the results presented in Fig. \eqref{Fig6} reinforce the consistency of the BPS regime adopted in this work and demonstrate how the generalized terms of the model affect the geometric and energetic organization of the vortex solutions, as well as the emergence of ring-like vortex configurations.
\begin{figure}[!ht]
    \centering
    \subfigure[Profile of the BPS energy density vs. $r$.]{\includegraphics[width=8cm,height=6cm]{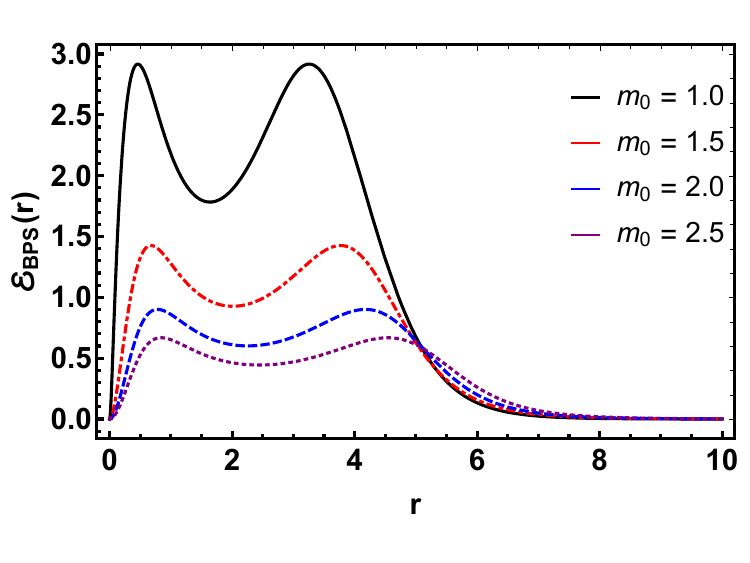}}\\
    \subfigure[The case $m_0=1.0$]{\includegraphics[width=4cm,height=4cm]{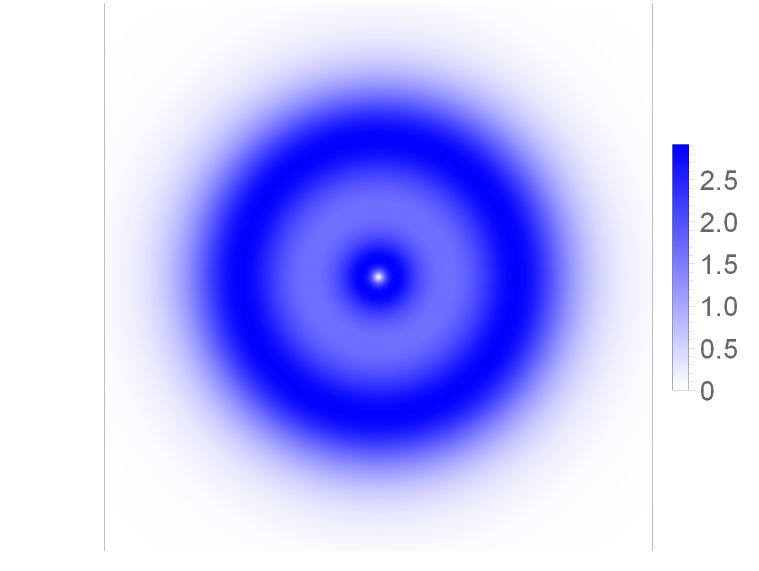}}\hfill
    \subfigure[The case $m_0=1.5$]{\includegraphics[width=4cm,height=4cm]{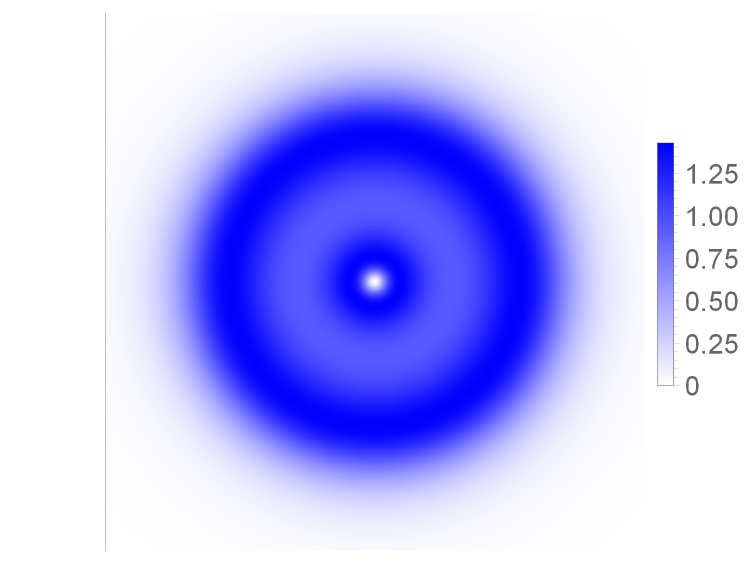}}\hfill
    \subfigure[The case $m_0=2.0$]{\includegraphics[width=4cm,height=4cm]{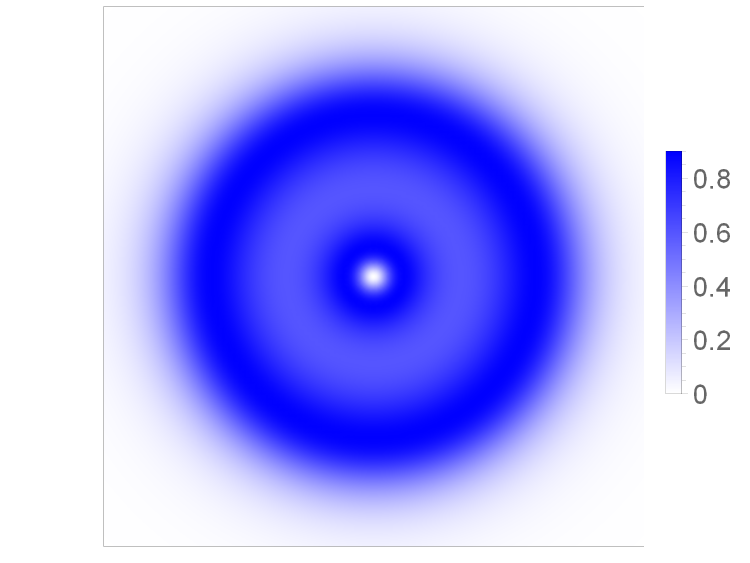}}\hfill
    \subfigure[The case $m_0=2.5$]{\includegraphics[width=4cm,height=4cm]{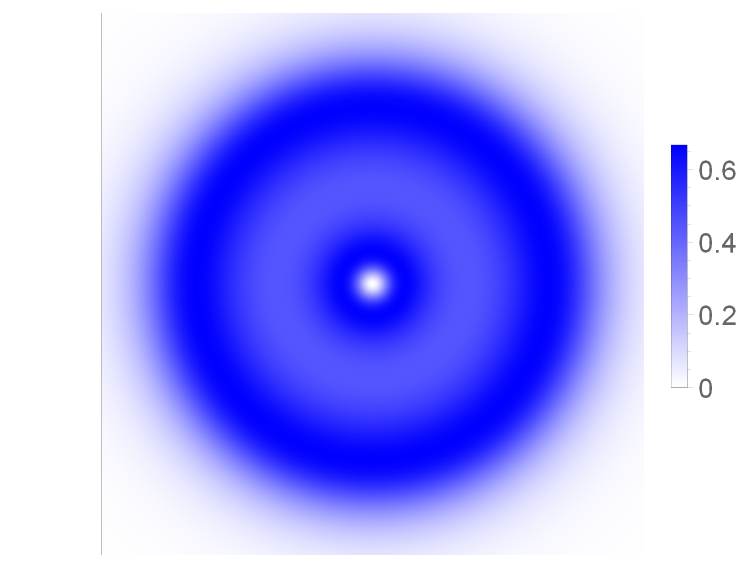}}
    \caption{Numerical solutions for the BPS energy density.}
    \label{Fig6}
\end{figure}

\subsection{Example 3: The case polynomial: $m(f)=\left\vert\mu\mathrm{e}^{\frac{16\pi^2P(f)^{-1}}{e^2}}\right\vert$}

To conclude, let us consider a polynomial function $P(f)$ such that the model effectively reduces to a Maxwell-Higgs-like theory, namely,
\begin{align}
    P(f)=f^{-2}(1+f^2)^2(1-f^2)^2.
\end{align}
Naturally, this choice reduces to the BPS equations in Eq. \eqref{Eq38} to
\begin{align}\label{Eq57}
    f'=\pm\frac{af}{r} \qquad \mathrm{and} \qquad \frac{a'}{r}=\mp 2 (1-f^2)^2.
\end{align}

\begin{figure}[!ht]
    \centering
    \includegraphics[width=8cm,height=7cm]{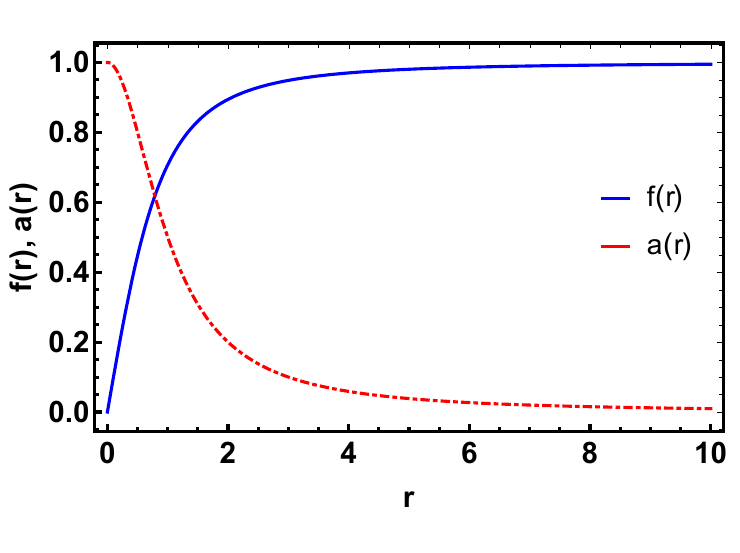}
    \caption{Numerical solutions of the variable fields $f(r)$ and $a(r)$ vs. $r$.}
    \label{Fig7}
\end{figure}

Taking into account the boundary conditions given in Eq. \eqref{Eq36}, we solve the BPS equations \eqref{Eq57} numerically. The corresponding numerical solutions of Eq. \eqref{Eq57} are displayed in Fig. \ref{Fig7}.

Upon examining Fig. \ref{Fig7}, one observes that the profile of the field variable $f(r)$ increases monotonically from the origin, satisfying the regularity condition $f(0)=0$. This behavior is consistent with the asymptotic analysis near the core, where $f(r)\sim r^{|n|}$, ensuring that the kinetic contributions associated with the angular term remain finite in the limit $r\to 0$. As $r$ increases, the field smoothly approaches the vacuum value $f(\infty)=1$, characterizing the typical profile of topological vortices. This behavior indicates that the $\CP^1$ field continuously interpolates between the defect core and the vacuum state in the asymptotic regime.

\begin{figure}[!ht]
    \centering
    \subfigure[Numerical solutions: $B(r)$ vs. $r$.]{\includegraphics[width=7 cm,height=6.5cm]{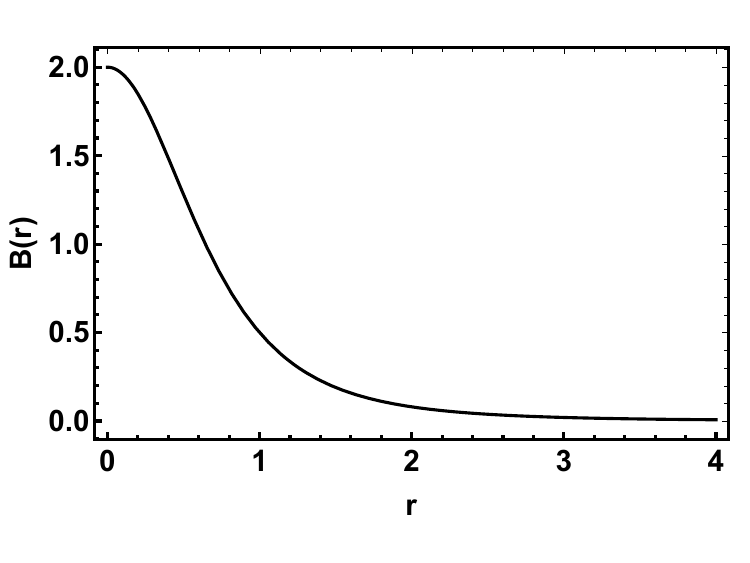}}
    \subfigure[Planar magnetic configuration.]{\includegraphics[width=5.5cm,height=5.5cm]{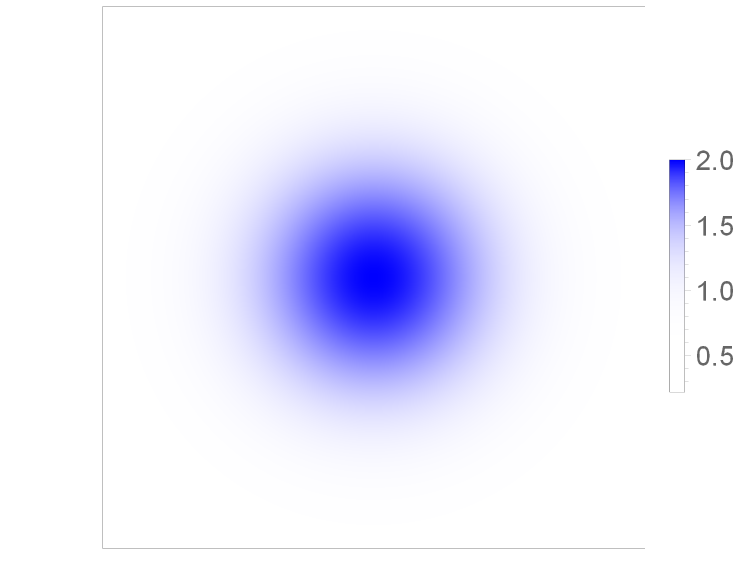}}
    \caption{Magnetic field $B(r)$ vs. $r$.}
    \label{Fig8}
\end{figure}

Meanwhile, the profile of the gauge function $a(r)$ starts from an integer value at the origin, determined by the topological winding number, and decreases monotonically until it vanishes as $r\to \infty$. This property ensures the compensation of the angular phase of the scalar field in the ansatz $u=f(r)\mathrm{e}^{in\theta}$, guaranteeing the regularity of the covariant derivative and the quantization of the magnetic flux. Consequently, the magnetic field associated with the configuration remains confined within a finite region around $r=0$ (see Fig. \ref{Fig8}), characterizing the topological structure as a magnetic vortex. Therefore, one concludes that the magnetic permeability, in this case, induces an effective dynamics capable of reproducing the typical properties of Maxwell-Higgs-like vortices.

\begin{figure}[!ht]
    \centering
    \subfigure[Plot of $\mathcal{E}_{\mathrm{BPS}}(r)$ vs. $r$]{\includegraphics[width=8cm,height=5.5cm]{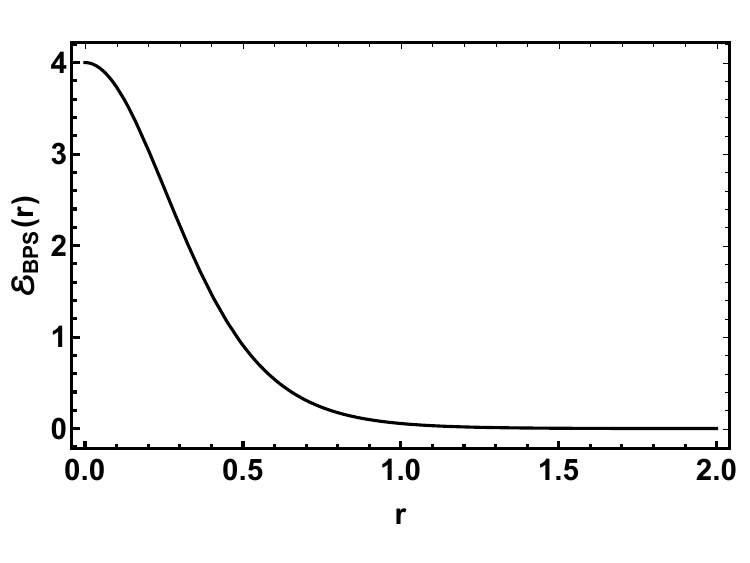}}
    \subfigure[Planar profile for BPS energy density.]{\includegraphics[width=5.5cm,height=5.5cm]{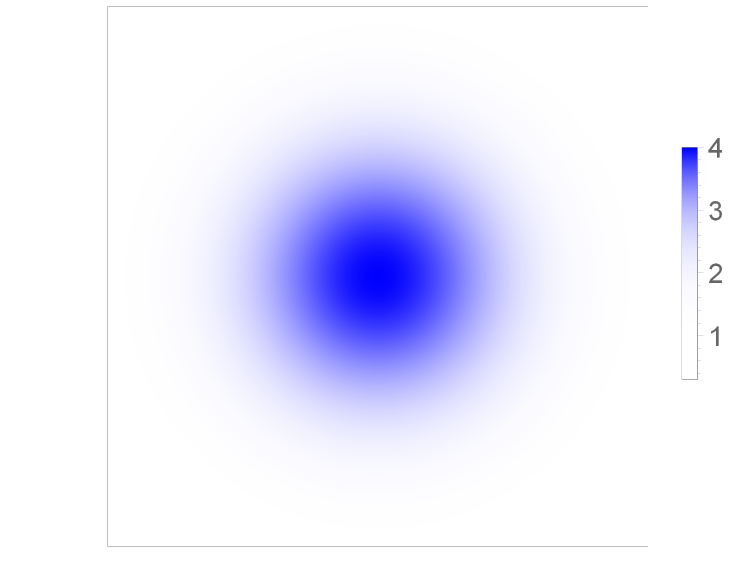}}
    \caption{BPS energy density $ \mathcal{E}_{\mathrm{BPS}}(r)$ vs. $r$ with unit winding number.}
    \label{Fig9}
\end{figure}

Finally, we examine the BPS energy density (Fig.~\ref{Fig9}), where one notes that the distributions exhibit a well-defined axial symmetry, directly reflecting the Nielsen-Olesen ansatz \cite{Nielsen}. Moreover, the maximum intensity remains concentrated near the vortex core and rapidly decreases as $r$ increases. Therefore, the spatial profiles confirm that both the energy and the magnetic flux are confined within a finite region of the plane. Furthermore, this behavior follows directly from the self-dual equation governing the gauge sector, $B=-a'(r)/r\propto(1-f^2)^2$. Since the profile $f(r)$ rapidly approaches the vacuum value $f\to 1$ in the asymptotic regime, the term $(1-f^2)^2$ tends to zero, leading to the suppression of both the magnetic field and the energy density far from the vortex core.
Thus, the compact circular structure exposed in Figs.~\ref{Fig8} and \ref{Fig9} indicates that the system supports energetically localized configurations, in which the total energy remains confined within a finite spatial region. This feature is essential to ensure that the total energy of the configuration is finite and proportional to the topological condition.

\section{Summary and conclusion}

In this work, we have investigated a generalized gauged $\CP^{1}$–Maxwell theory in which the electromagnetic sector acquires a field–dependent magnetic permeability dynamically generated through fermionic vacuum polarization. Starting from a microscopic framework in which a Dirac fermion is minimally coupled to the gauge field and endowed with an effective mass depending on the $\CP^{1}$ scalar sector, we have shown that quantum fluctuations at the one–loop level induce a non–polynomial modification of the Maxwell term. In particular, the vacuum polarization tensor generates an effective dielectric function whose logarithmic structure arises naturally from the renormalized fermionic determinant. After performing a dimensional reduction from $(3+1)$ to $(2+1)$ dimensions, this mechanism yields a generalized Maxwell sector characterized by a magnetic permeability $G(\vert u\vert)$ explicitly controlled by the dynamics of the $\CP^{1}$ field.

Within this effective framework, we constructed a generalized $\CP^{1}$–Maxwell model that supports the BPS property. By employing the Bogomol'nyi procedure, the energy functional was rewritten as a sum of positive-definite terms plus a topological contribution, allowing the derivation of first-order self-dual (or BPS) equations. A crucial aspect of this construction is the nontrivial interplay between the geometric structure of the target space, encoded in the Fubini-Study metric, and the dielectric function governing the electromagnetic sector. The presence of these two ingredients significantly modifies the dynamical structure. In particular, the scalar field evolves on a curved manifold while simultaneously determining the effective electromagnetic medium through which the gauge field propagates. 

The BPS equations derived in this framework describe magnetized vortex-like configurations with quantized magnetic flux. By adopting a Nielsen-Olesen-like radial ansatz, we reduced the system to a pair of self-dual equations that determine the radial profiles of the scalar and gauge fields. The resulting solutions exhibit the characteristic properties of BPS vortices, i.e., regularity at the origin, finite energy, and magnetic flux quantization. Moreover, the asymptotic analysis revealed that the scalar and gauge fluctuations decay exponentially, with a characteristic mass scale that depends explicitly on the magnetic permeability evaluated at the vacuum configuration. This result demonstrates that the induced dielectric function controls the intrinsic length scale of the vortex solutions, thereby establishing a direct link between the effective vacuum structure and the spatial width of the topological defects.

A detailed analysis of the asymptotic regimes confirmed the regularity and stability of the magnetized configurations. Near the origin, the scalar field vanishes with a power-law behavior determined by the winding number, ensuring the finiteness of the angular kinetic term and preventing singularities in the energy density. At spatial infinity, the solutions approach the vacuum configuration in a manner dictated by the effective mass of the coupled scalar-gauge mode. In the topological sector, this leads to exponentially localized configurations whose magnetic flux and energy density remain confined within a finite spatial region. In contrast, in regimes where the effective fermionic mass approaches zero, the dielectric function diverges logarithmically, producing long-range configurations characterized by power-law asymptotic behavior.

To explore these features more explicitly, we investigated several representative examples corresponding to different choices of the effective fermionic mass profile. The first example reproduces the standard $\CP^{1}$-Maxwell theory with constant magnetic permeability, serving as a reference configuration that recovers previously known BPS structures. In this limit, the solutions correspond to magnetized lump-like configurations whose magnetic field and energy density remain localized around the core. The second example introduces a logarithmic dielectric function arising from a polynomial dependence of the fermionic mass on the scalar field. In this case, the magnetic permeability modifies the coupling between the scalar and magnetic sectors, resulting in the deformation of the lump-like profiles to magnetic vortex configurations. Meanwhile, in the third example, an appropriate choice of the mass function leads to a polynomial dielectric structure that effectively reproduces vortex configurations, demonstrating that the generalized framework is capable of interpolating between different classes of topological defects.

From a broader perspective, the present work reveals how quantum effects originating in the fermionic sector can generate nontrivial modifications in the effective gauge dynamics, which in turn influence the properties of topological solitons. The induced magnetic permeability acts as a dynamical medium through which the gauge field propagates, introducing an additional layer of structure into the vortex solutions. In this sense, the model provides a concrete realization of how radiative corrections can reshape the nonperturbative sector of gauge theories.

In summary, the generalized $\CP^{1}$-Maxwell theory provides a consistent framework in which vacuum polarization effects dynamically generate a field-dependent electromagnetic permeability capable of supporting two-dimensional BPS configurations. The resulting solutions exhibit quantized magnetic flux, finite energy, and localized energy densities whose spatial extension is controlled by the magnetic permeability. 

These results open several directions for future investigation, including the analysis of moduli-space dynamics, the study of multi-vortex interactions, and possible extensions involving supersymmetric embeddings or Chern-Simons terms. Such developments may further clarify the role played by quantum-induced electromagnetic media in the formation and dynamics of topological defects in gauge theories.

\section*{Acknowledgment}

This work was supported by the São Paulo Research Foundation (FAPESP) under Grant No. 2025/05176-7 (FCEL).

\bibliographystyle{apsrev4-2}
\bibliography{refs}

@article{Bogomolnyi,
  title={The stability of classical solutions},
  author={Bogomol'Nyi, EB},
  journal={Sov. J. Nucl. Phys.(Engl. Transl.);(United States)},
  volume={24},
  number={4},
  year={1976},
  publisher={LD Landau Theoretical Physics Institute, USSR Academy of Sciences, Moscow}
}

@article{Prasad,
  title={Exact classical solution for the't Hooft monopole and the Julia-Zee dyon},
  author={Prasad, Manoj K and Sommerfield, Charles M},
  journal={Phys. Rev. Lett.},
  volume={35},
  number={12},
  pages={760},
  year={1975},
  publisher={APS}
}

@article{Adam,
  title={The first-order Euler-Lagrange equations and some of their uses},
  author={Adam, C and Santamaria, F},
  journal={JHEP},
  volume={2016},
  number={12},
  pages={47},
  year={2016},
  publisher={Springer}
}

@article{Atmaja,
  title={A method for BPS equations of vortices},
  author={Atmaja, A Nata},
  journal={Phys. Lett. B},
  volume={768},
  pages={351--358},
  year={2017},
  publisher={Elsevier}
}

@article{Tong,
  title={Moduli space of BPS domain walls},
  author={Tong, David},
  journal={Phys. Rev. D},
  volume={66},
  number={2},
  pages={025013},
  year={2002},
  publisher={APS}
}

@article{Gibbons,
  title={The moduli space metric for well-separated BPS monopoles},
  author={Gibbons, Gary W and Manton, Nicholas S},
  journal={Phys. Lett. B},
  volume={356},
  number={1},
  pages={32--38},
  year={1995},
  publisher={Elsevier}
}

@article{Lee,
  title={Moduli space of many BPS monopoles for arbitrary gauge groups},
  author={Lee, Kimyeong and Weinberg, Erick J and Yi, Piljin},
  journal={Phys. Rev. D},
  volume={54},
  number={2},
  pages={1633},
  year={1996},
  publisher={APS}
}

@article{Witten,
  title={Supersymmetry algebras that include topological charges},
  author={Witten, Edward and Olive, David},
  journal={Phys. Lett. B},
  volume={78},
  number={1},
  pages={97--101},
  year={1978},
  publisher={Elsevier}
}

@incollection{Intriligator,
  title={Lectures on supersymmetric gauge theories and electric-magnetic duality},
  author={Intriligator, K and Seiberg, N},
  booktitle={Low-Dimensional Applications of Quantum Field Theory},
  pages={161--199},
  year={1997},
  publisher={Springer}
}

@article{Manton,
  title={\emph{A $\CP^2$ SMEFT}},
  author={Manton, NS},
  journal={JHEP},
  volume={2025},
  number={4},
  pages={1--18},
  year={2025},
  publisher={Springer}
}

@article{Bazeia1,
  title={Generalized Maxwell--Higgs vortices in models with enhanced symmetry},
  author={Bazeia, D and Liao, MA and Marques, MA},
  journal={Eur. Phys. J. C},
  volume={82},
  number={4},
  pages={316},
  year={2022},
  publisher={Springer}
}

@article{Lima1,
  title={Ring-like vortices in a logarithmic generalized Maxwell theory},
  author={Lima, F C E and Almeida, C A S},
  journal={Europhys. Lett.},
  volume={131},
  number={3},
  pages={31003},
  year={2020},
  publisher={EDP Sciences, IOP Publishing and Societ{\`a} Italiana di Fisica}
}

@article{Andrade,
  title={Vortices in Maxwell-Higgs models with a global factor},
  author={Andrade, I and Marques, M A and Menezes, R},
  journal={Europhys. Lett.},
  volume={133},
  number={3},
  pages={31002},
  year={2021},
  publisher={EDP Sciences, IOP Publishing and Societ{\`a} Italiana di Fisica}
}

@article{Ward,
  title={Slowly-moving lumps in the CP1 model in (2+ 1) dimensions},
  author={Ward, RS},
  journal={Phys. Lett. B},
  volume={158},
  number={5},
  pages={424--428},
  year={1985},
  publisher={Elsevier}
}

@article{Leese,
  title={Low-energy scattering of solitons in the CP1 model},
  author={Leese, Robert},
  journal={Nuc. Phys. B},
  volume={344},
  number={1},
  pages={33--72},
  year={1990},
  publisher={Elsevier}
}

@article{Nahum,
  title={3d loop models and the cp n-1 sigma model},
  author={Nahum, Adam and Chalker, JT and Serna, P and Ortuno, M and Somoza, AM},
  journal={Phys. Rev. Lett.},
  volume={107},
  number={11},
  pages={110601},
  year={2011},
  publisher={APS}
}

@article{Buccio,
  title={Renormalization and running in the 2D CP (1) model},
  author={Buccio, Diego and Donoghue, John F and Menezes, Gabriel and Percacci, Roberto},
  journal={JHEP},
  volume={2025},
  number={2},
  pages={1--15},
  year={2025},
  publisher={Springer}
}

@book{Polyakov,
  title={Gauge fields and strings},
  author={Polyakov, Aleksandr Michajlovi{\v{c}}},
  year={2018},
  publisher={Routledge}
}

@article{Belavin,
  title={Metastable states of two-dimensional isotropic ferromagnets},
  author={Belavin, AA and Polyakov, AM},
  journal={JETP lett},
  volume={22},
  number={10},
  pages={245--248},
  year={1975}
}

@article{Lian,
  title={Small instantons in CP 1 and CP 2 sigma models},
  author={Lian, Yaogang and Thacker, HB},
  journal={Phys. Rev. D},
  volume={75},
  number={6},
  pages={065031},
  year={2007},
  publisher={APS}
}

@book{Rajaraman,
  title={Solitons and instantons. An introduction to solitons and instantons in quantum field theory},
  author={Rajaraman, Ramamurti},
  year={1982}
}

@article{Bazeia2,
  title={Planar ringlike vortices},
  author={Bazeia, D and Marques, MA and Melnikov, D},
  journal={Phys. Lett. B},
  volume={785},
  pages={454--461},
  year={2018},
  publisher={Elsevier}
}

@article{Bazeia3,
  title={Compact vortices},
  author={Bazeia, D and Losano, L and Marques, MA and Menezes, R and Zafalan, I},
  journal={Eur. Phys. J. C},
  volume={77},
  number={2},
  pages={63},
  year={2017},
  publisher={Springer}
}

@article{Lima2,
  title={Exponentially generalized vortex},
  author={Lima, F C E and Almeida, C A S},
  journal={Europhys. Lett.},
  volume={138},
  number={4},
  pages={44001},
  year={2022},
  publisher={EDP Sciences, IOP Publishing and Societ{\`a} Italiana di Fisica}
}

@article{Lima3,
  title={Ring-like vortices in a logarithmic generalized Maxwell theory},
  author={Lima, F C E and Almeida, C A S},
  journal={Europhys. Lett.},
  volume={131},
  number={3},
  pages={31003},
  year={2020},
  publisher={EDP Sciences, IOP Publishing and Societ{\`a} Italiana di Fisica}
}

@article{Petrov,
  title={Vortex solutions in nonpolynomial scalar QED},
  author={Lima, F C E and Petrov, A Yu and Almeida, C A S},
  journal={Phys. Rev. D},
  volume={103},
  number={9},
  pages={096019},
  year={2021},
  publisher={APS}
}

@article{Jackiw,
  title={Radiatively induced Lorentz and CPT violation in electrodynamics},
  author={Jackiw, Roman and Kosteleck{\`y}, V Alan},
  journal={Phys. Rev. Lett.},
  volume={82},
  number={18},
  pages={3572},
  year={1999},
  publisher={APS}
}

@article{Nielsen,
  title={Vortex-line models for dual strings},
  author={Nielsen, Holger Bech and Olesen, Poul},
  journal={Nucl. Phys. B},
  volume={61},
  pages={45--61},
  year={1973},
  publisher={Elsevier}
}

@book{VachaspatiB,
  title={Kinks and domain walls: An introduction to classical and quantum solitons},
  author={Vachaspati, Tanmay},
  year={2007},
  publisher={Cambridge University Press}
}

@article{Jackiw00,
  title={Self-dual Chern-Simons solitons},
  author={Jackiw, R and Lee, Kimyeong and Weinberg, Erick J},
  journal={Phys. Rev. D},
  volume={42},
  number={10},
  pages={3488},
  year={1990},
  publisher={APS}
}

@book{Manton2004,
  title={Topological solitons},
  author={Manton, Nicholas and Sutcliffe, Paul},
  year={2004},
  publisher={Cambridge University Press}
}

@article{Cunha,
  title={Magnetized BPS lumps in the $ CP\^{} 1$ model with Maxwell coupling},
  author={Cunha, I B and Lima, F C E and Vera, Aldo},
  journal={arXiv preprint arXiv:2602.22957},
  year={2026}
}

@article{Abrikosov1,
  title={On the magnetic properties of superconductors of the second group},
  author={Abrikosov, Alexei A},
  journal={Soviet Physics-JETP},
  volume={5},
  pages={1174--1182},
  year={1957}
}

@article{Abrikosov2,
  title={On the magnetic properties os superconductors of the second group},
  author={Abrikosov, AA},
  journal={Sov. Phys. J. Exp. Theor. Phys.},
  volume={32},
  pages={1442},
  year={1955}
}

@incollection{Ginzburg,
  title={On the theory of superconductivity},
  author={Ginzburg, Vitaly L and Landau, Lev D},
  booktitle={On superconductivity and superfluidity: a scientific autobiography},
  pages={113--137},
  year={2009},
  publisher={Springer}
}

@article{Kibble1,
  title={Symmetry breaking in non-Abelian gauge theories},
  author={Kibble, Tom WB},
  journal={Phys. Rev.},
  volume={155},
  number={5},
  pages={1554},
  year={1967},
  publisher={APS}
}

@article{Kibble2,
  title={Topology of cosmic domains and strings},
  author={Kibble, Thomas WB},
  journal={Journal of Physics A: Mathematical and General},
  volume={9},
  number={8},
  pages={1387--1398},
  year={1976}
}

@book{Vilenkin,
  title={Cosmic strings and other topological defects},
  author={Vilenkin, Alexander and Vilenkin, Alexander and Shellard, EPS},
  year={1994},
  publisher={Cambridge University Press}
}

@article{Linet,
  title={A vortex-line model for infinite straight cosmic strings},
  author={Linet, B},
  journal={Phys. Lett. A},
  volume={124},
  number={4-5},
  pages={240--242},
  year={1987},
  publisher={Elsevier}
}

@article{Kim,
  title={Global vortex and black cosmic string},
  author={Kim, Nakwoo and Kim, Yoonbai and Kimm, Kyoungtae},
  journal={Phys. Rev. D},
  volume={56},
  number={12},
  pages={8029},
  year={1997},
  publisher={APS}
}

@article{Fedderke,
  title={Periodic cosmic string formation and dynamics},
  author={Fedderke, Michael A and Huang, Junwu and Siemonsen, Nils},
  journal={JHEP},
  volume={2025},
  number={8},
  pages={1--77},
  year={2025},
  publisher={Springer}
}

@article{Hayata,
  title={Phase transition on superfluid vortices in Higgs-Confinement crossover},
  author={Hayata, Tomoya and Hidaka, Yoshimasa and Kondo, Dan},
  journal={JHEP},
  volume={2025},
  number={3},
  pages={1--21},
  year={2025},
  publisher={Springer}
}

@article{Martinec,
  title={BPS fivebrane stars. Part III. Effective actions},
  author={Martinec, Emil J and Zigdon, Yoav},
  journal={JHEP},
  volume={2025},
  number={3},
  pages={1--46},
  year={2025},
  publisher={Springer}
}

@article{Alonso2,
  title={Dynamics of excited BPS three-vortices},
  author={Alonso Izquierdo, A and Manton, NS and Mateos Guilarte, J and Rees, M and Wereszczynski, A},
  journal={Phys. Rev. D},
  volume={111},
  number={10},
  pages={105021},
  year={2025},
  publisher={APS}
}

@article{Lu2025,
  title={New BPS states from bosonic/heterotic duality},
  author={Lu, Kai-Peng and L{\"u}, H and Ma, Liang},
  journal={JHEP},
  volume={2025},
  number={7},
  pages={1--20},
  year={2025},
  publisher={Springer}
}

@article{Kim2025,
  title={Inhomogeneous abelian Chern-Simons Higgs model with new inhomogeneous BPS vacuum and solitons},
  author={Kim, Yoonbai and Kwon, O-Kab and Song, Hanwool and Kim, Chanju},
  journal={JHEP},
  volume={2025},
  number={3},
  pages={1--25},
  year={2025},
  publisher={Springer}
}

@article{Andrade22,
  title={BPS chiral vortices in Maxwell-Higgs electrodynamics},
  author={Andrade, J and Casana, Rodolfo and da Hora, E},
  journal={Phys. Rev. D},
  volume={111},
  number={3},
  pages={036019},
  year={2025},
  publisher={APS}
}

@article{IAndrade11,
  title={First-order framework for vortices in generalized Maxwell--Chern--Simons models without a neutral field},
  author={Andrade, I and Bazeia, D and Liao, MA and Marques, MA and Menezes, R},
  journal={Mod. Phys. Lett. A},
  volume={37},
  number={33n34},
  pages={2250225},
  year={2022},
  publisher={World Scientific}
}

@article{Weinberg1968,
  title={Nonlinear realizations of chiral symmetry},
  author={Weinberg, Steven},
  journal={Phys. Rev.},
  volume={166},
  number={5},
  pages={1568},
  year={1968},
  publisher={APS}
}

@article{Coleman1969,
  title={Structure of phenomenological Lagrangians. I},
  author={Coleman, Sidney and Wess, Julius and Zumino, Bruno},
  journal={Phys. Rev.},
  volume={177},
  number={5},
  pages={2239},
  year={1969},
  publisher={APS}
}

@article{Callan1969,
  title={Structure of phenomenological Lagrangians. II},
  author={Callan Jr, Curtis G and Coleman, Sidney and Wess, Julius and Zumino, Bruno},
  journal={Phys. Rev.},
  volume={177},
  number={5},
  pages={2247},
  year={1969},
  publisher={APS}
}

@article{Polyakov2,
  title={Interaction of goldstone particles in two dimensions. Applications to ferromagnets and massive Yang-Mills fields},
  author={Polyakov, Alexander M},
  journal={Phys. Lett. B},
  volume={59},
  number={1},
  pages={79--81},
  year={1975},
  publisher={Elsevier}
}

@article{Manton2,
  title={A $$$\{$$\backslash$mathbbm $\{$CP$\}$$\}$\^{} 2$$ SMEFT},
  author={Manton, NS},
  journal={Journal of High Energy Physics},
  volume={2025},
  number={4},
  pages={1--18},
  year={2025},
  publisher={Springer}
}

@article{RevAtmaja,
  title={Bogomol’nyi equations of classical solutions},
  author={Atmaja, Ardian N and Ramadhan, Handhika S},
  journal={Phys. Rev. D},
  volume={90},
  number={10},
  pages={105009},
  year={2014},
  publisher={APS}
}

@article{RevRamadhan,
  title={Bogomol’nyi equations for Dirac--Born--Infeld cosmic string},
  author={Ramadhan, Handhika Satrio and Athaullah, M Naufal and Prasetyo, Ilham},
  journal={The European Physical Journal C},
  volume={85},
  number={12},
  pages={1452},
  year={2025},
  publisher={Springer}
}

\end{document}